\documentclass[sigconf]{acmart} 

\AtBeginDocument{%
  }

\setcopyright{acmlicensed}
\copyrightyear{2025}
\acmYear{2025}
\acmDOI{XXXXXXX.XXXXXXX}

\acmConference[AM '25]{Audio Mostly 2025}{June 30--July 04,
  2025}{Coimbra, Portugal}
\acmISBN{978-1-4503-XXXX-X/18/06}

\usepackage{subfiles}
\usepackage{cleveref}
\usepackage{booktabs}
\usepackage{longtable}
\newcommand{\updownarrows}{\uparrow\mathrel{\mspace{-1mu}}\downarrow}
\newcommand{\downuparrows}{\downarrow\mathrel{\mspace{-1mu}}\uparrow}
\renewcommand{\upuparrows}{\uparrow\uparrow}
\renewcommand{\downdownarrows}{\downarrow\downarrow}
\newcommand{\inlinemaths}[2][1.8mm]{\raisebox{0.4mm}{\resizebox{#1}{1.8mm}{$#2$}}}
\DeclareUnicodeCharacter{21C8}{\inlinemaths{\upuparrows}}

\DeclareUnicodeCharacter{21C5}{\inlinemaths{\updownarrows}}

\DeclareUnicodeCharacter{21F5}{\inlinemaths{\downuparrows}}

\DeclareUnicodeCharacter{21CA}{\inlinemaths{\downdownarrows}}

\DeclareUnicodeCharacter{2191}{\inlinemaths[1.2mm]{\,\uparrow}}
\newcommand{\polarup}{↑}
\DeclareUnicodeCharacter{2193}{\inlinemaths[1.2mm]{\,\downarrow}}
\newcommand{\polardown}{↓}

\newcommand{\embraced}[1]{\inlinemaths{\langle}#1\inlinemaths{\rangle}}

\begin{document}

%% Paper title.
\title{Two Empirical Studies on Audiovisual Semiotics of Uncertainty}

\author{Sita A. Vriend}
\email{sita.vriend@visus.uni-stuttgart.de}
\orcid{0009-0008-8530-835X}
\affiliation{%
  \institution{University of Stuttgart}
  \city{Stuttgart}
  \country{Germany}
}

\author{David H\"{a}gele}
\email{david.haegele@visus.uni-stuttgart.de}
\orcid{0000-0002-2679-6882}
\affiliation{%
  \institution{University of Stuttgart}
  \city{Stuttgart}
  \country{Germany}
}

\author{Daniel Weiskopf}
\email{daniel.weiskopf@visus.uni-stuttgart.de}
\orcid{0000-0003-1174-1026}
\affiliation{%
  \institution{University of Stuttgart}
  \city{Stuttgart}
  \country{Germany}
}

%\renewcommand{\shortauthors}{Vriend et al.}

\begin{abstract}
There exists limited theoretical guidance on integrating visualization and sonification. In this paper, we address this gap by investigating audiovisual semiotics for uncertainty representation: joining uncertainty visualization and sonification to combine audiovisual channels for enhancing users' perception of uncertainty. We conducted two preregistered crowd-sourced user studies. First, we assessed suitable audio/visual pairs. Then, we investigated audiovisual mappings of uncertainty. Here, we use probability as it is an easily communicated aspect of uncertainty. We analyzed the participants' preferences and reaction times in both user studies. Additionally, we explored the strategies employed by participants through qualitative analysis. Our results reveal audiovisual mappings that lead to particularly strong preferences and low reaction times. Furthermore, we found that preferred audio/visual pairs are not necessarily suitable audiovisual mappings of uncertainty. For example, while pitch paired with brightness was preferred as a pair, it was not well suited as a mapping for uncertainty. We recommend audiovisual mappings of uncertainty that lead to low reaction times and high preferences in both user studies. This paper presents guidelines to anyone seeking to employ audiovisual representations for uncertainty, contributing to enhancing the perception of uncertainty.

\end{abstract}



\begin{teaserfigure}
  \includegraphics[width=\textwidth]{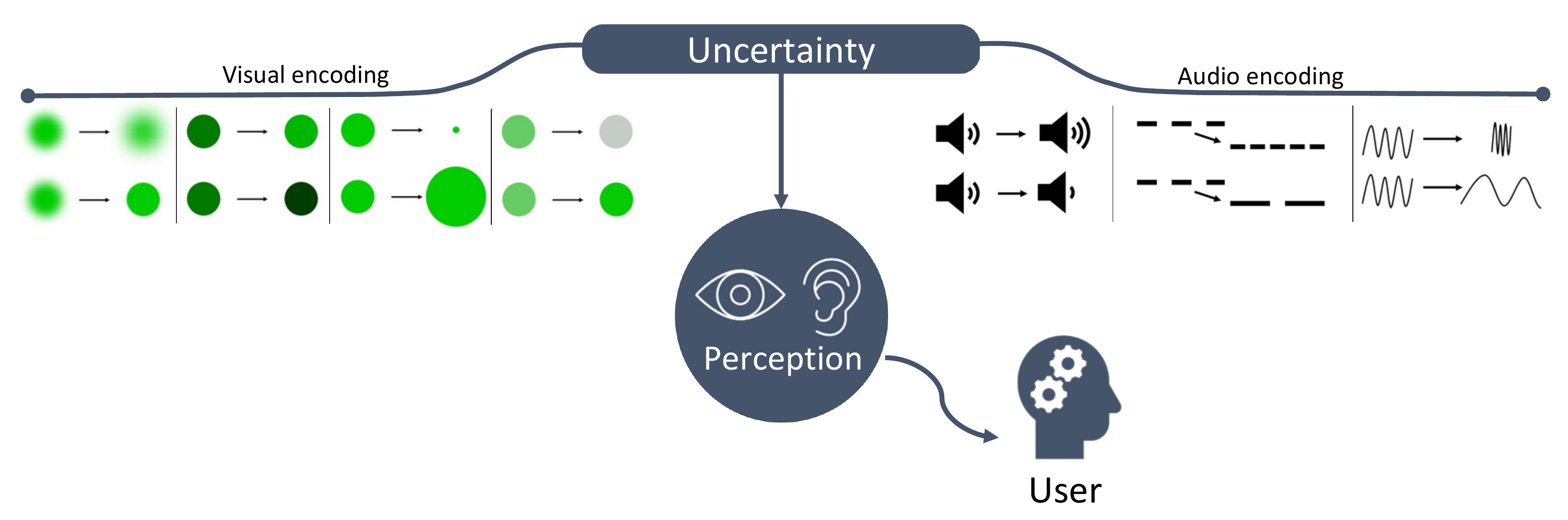}
  \caption{The conceptual process of audiovisual semiotics of uncertainty. Uncertainty is encoded in the audio and visual channels simultaneously, forming an audiovisual pair.
  The sound dimensions loudness, tempo, and pitch pair with fuzziness, brightness, size, and saturation.
  The individual combinations and their polarity for mapping uncertainty are perceived differently by the user, and their preference varies strongly.}
  \Description{The figure illustrates the relationship between audio and visual encodings and user perception of uncertainty. The central theme of the figure is Uncertainty, placed prominently. On either side of this term, there are two elements: Audio Encoding and Visual Encoding. Both these encodings appear to interact with the User Perception at the bottom of the figure. The layout visually emphasizes how uncertainty is involved in how users perceive stimuli encoded through both audio and visual means.}
  \label{fig:teaser}
\end{teaserfigure}

%\received{12 September 2024}
%\received[revised]{20 March 2025}
\received[accepted]{5 May 2025}

\maketitle

%%%%%%%%%%%%%% Introduction %%%%%%%%%%%%%%%%%%%%

\section{Introduction}

% uncertainty - is hard to understand
Uncertainty can arise at any stage of the analysis pipeline. It is valuable to propagate and communicate this uncertainty to the user. In science, we often quantify and report uncertainty alongside experimental results, to indicate the level of confidence or error of the measurements. Outside of academia, fields like weather forecasting~\cite{witt2023visualizing} and healthcare \cite{fagerlin2005reducing} account for uncertainty, often in the form of probabilities. Uncertainty visualization has been studied extensively~\cite{yang2024trust, Weiskopf2022Uncertainty, haegeleIt2022, Brodlie:2012:RUD, Bonneau:2014:OSU}. However, uncertainty remains difficult to understand even when visualized~\cite{hoekstra2014robust}, even for experts~\cite{belia2005researchers}. 

% Semiotics approach visual semiotics
One approach to visualizing uncertainty is to use perceptual channels that align with the user's understanding of the data \cite{maceachren2012visual, bertin1983semiology}. Audio is a popular alternative modality for data representation, relying on the process of sonification \cite{kim2024erie}. Sonification can easily be used together with visualization.  However, limited research has integrated sonification and visualization \cite{enge2023towards, enge2024open}, leaving researchers with little guidance for design decisions. This includes determining suitable audiovisual (AV) mappings for concepts like uncertainty.

% sonification/visualization choices
Sonification design choices \cite{walker2002magnitude, walker2005mappings} are similar to those of visualization \cite{munzner2014visualization, correll2018value}. While visualization researchers consider how to represent data using visual channels and marks, sonification researchers decide on the sound channel to represent a given data channel. Second, both visualization and sonification researchers consider how much change in the visual or audio channel represents a change in the given variable, i.e., scaling. Third, visualization and sonification consider polarity: whether the direction in the sound or visual channel should increase or decrease to represent the variable's increase. \cite{hermann2011sonification}. However, the polarity and sound channel choice is especially challenging in sonification \cite{walker2002magnitude, walker2005mappings}. Polarity preference differs between different data variables \cite{hermann2011sonification}. For the same data variable, participants often have different polarity preferences \cite{dubus2013systematic, walker2002magnitude}.

% AV analytics design choices
Combining visualization and sonification leads to additional design choices: \textbf{1)~Which audio and visual channel to pair.} Some visual channels, such as \embraced{loudness/size}, seem to pair more naturally with certain audio channels compared to others \cite{parise2013audiovisual}. \textbf{ 2)~The polarity matching of the AV pairs}. For some pairs, like \embraced{pitch/brightness}, users prefer the polarity of both channels to increase. In contrast, some pairs, such as \embraced{loudness/fuzziness}, might be preferred if the polarities do not match. \textbf{3)~The mapping between the AV channel and the data concept} is important. Just because \embraced{pitch\polarup \polarup brightness} is preferred, in this case with matching polarities, does not mean that the AV pair is a suitable representation of uncertainty.  
\Cref{fig:teaser} illustrates the pairing of AV channels for conveying uncertainty.

% set up paper
Inspired by earlier research on visual semiotics \cite{maceachren2012visual} and uncertainty sonification \cite{ballatore2019sonifying}, we investigate AV semiotics of uncertainty with two preregistered crowd-sourced user studies. Research into AV cross-modal correspondences \cite{parise2013audiovisual} has not yet investigated audio and visual channels typically used for uncertainty. Hence, in the first experiment, we ask \textbf{RQ1}: \textbf{Which pairs of audio and visual stimuli are most preferred?} However, preferred AV pairs are not automatically suitable uncertainty mappings. Therefore, our second research question \textbf{RQ2} asks: \textbf{Which audiovisual mappings are most preferred to represent uncertainty?} In a second experiment, we examine whether the AV pairs studied in Experiment~1 are suitable for representing uncertainty. We ask participants to map AV stimuli to probability since uncertainty itself is poorly understood by participants in previous research \cite{ballatore2019sonifying}. Furthermore, probability is a relatable aspect of uncertainty, making it easier for participants to interpret the stimuli. In this research, we focus specifically on AV pairs and AV mappings of probability and their preferred polarity. Hence, we measure the participants' polarity preferences and reaction times (RTs) in both experiments. Finally, to understand how users integrate AV semiotics to perceive uncertainty we asked what strategy participants employed during the~experiments. 

% Aim of this work + contribution 
While prior research often examines sonification or visualization in isolation, this research explores the integration of sonification and visualization  \cite{enge2023towards, enge2024open}. Furthermore, this paper provides empirical insights into uncertainty encodings. Finally, we provide design implications for integrating AV representations into uncertainty visualization systems. In summary, our results are instrumental in providing guidelines for designing AV data displays including uncertainty.

%%%%%%%%%%% Related work %%%%%%%%%%%%%%

\section{Related Work}
Sonification, according to Kramer et al. \cite{kramer2010sonification}, ``is the use of nonspeech audio to convey information.'' Visualization, in contrast, is defined by Munzer \cite{munzner2014visualization} as ``visual representations of datasets.''  

Visualization and sonification use the same approach to convey data~\cite{aigner2022workshop, enge2024open, enge2023towards}, thus their theories are largely compatible. Both encode data using perceptual variables to convey information to a user. In visualization, such variables can be visual channels, e.g., as color or size, while sonification uses audio channels, such as loudness or pitch. Due to these similarities, integrating them is~promising.

\subsection{Uncertainty Visualization}
% uncertainty
Quantified uncertainty is an umbrella term for a wide variety of measures, such as error or variability. 
In fact, uncertainty on its own is considered a vague term by non-experts \cite{ballatore2019sonifying}. While the inverse of uncertainty---certainty---can be relevant, fields that study uncertainty benefit from indicating uncertainties. Uncertainty visualization concerns itself with measures of uncertainty. It is a diverse topic involving fields beyond computer science concerned with uncertainty. Accuracy, precision, completeness, and credibility are relevant indicators of uncertainty in cartography and, thus, frequently visualized \cite{maceachren2012visual}. Probability is a common measure of uncertainty in everyday life. Probability of outcomes is relevant in research involving forecasts \cite{ruginski2016non}, which extends to everyday life to communicate weather predictions. Probability is also a common measure of uncertainty in health informatics to communicate risks of medical procedures to patients \cite{galesic2009using}. Because of its commonality, probability is a relatable and quantifiable form of uncertainty, making it easier for participants to interpret the stimuli. Hence, we asked participants to map the AV stimuli to probability in this research. 

% design space
The uncertainty visualization design space can be broken down into three approaches: graphical annotations of distributional properties, visual encodings of uncertainty, and hybrid approaches. Graphical annotations of distributional properties are typically used to visualize probability. For example, probability density plots map probability to height, and gradient plots map probability to opacity \cite{padilla2020uncertainty}. 

% visual encodings
In this research, we focus on encodings of uncertainty. Some visual encodings seem to be processed immediately and preconceptually at a sensory level \cite{roth2017visual}, which makes the data easier to interpret. Prior work \cite{maceachren2012visual, leitner2000guidelines} studied fuzziness, color, size, and saturation as visual semiotics of uncertainty. Our study extends this research by integrating sonification.

% hybrid
Hybrid approaches add visual encodings to graphical annotations of distributional properties~\cite{boukhelifa2012sketchiness}. Fuzziness positively affects participants' perception of uncertainty in temporal visualizations~\cite{gschwandtnei2016temporal} and line graphs~\cite{pinney2023aesthetic}. The AV encodings we explore can be combined with graphical annotations of distributional properties by future research to inspire new AV hybrid approaches.

\subsection{Sonification}
% benefits
Presenting data using audio has many benefits. Our auditory perception is particularly sensitive to temporal changes, recognizing patterns \cite{lotto2011psychology}, and detecting subtle auditory variations compared to visual cues \cite{kramer2010sonification}. Furthermore, we are familiar with information presented through sound. Modern cars, for example, use sound cues to inform drivers, such as reminding them to fasten their seatbelts and warning them of potential problems \cite{Walker2023promise}. 

% accessibility
Sonification can substitute visualization in the context of accessibility for blind people~\cite{cooke2017exploring, Tomlinson2020science, kimmarriott2022chi, bongshinlee2023chi, Hoque2023natural, niklas2024tvcg, Moore2024spatial}. While we expect that our results partly carry over to this context, our primary goal is not to substitute hearing for vision, but to combine both senses.

% semiotics channels and polarity explored
The concept of semiotics naturally aligns with sonification. Sonification researchers commonly select audio channels such as pitch, tempo, and loudness to encode data~\cite{dubus2013systematic}. Furthermore, research has investigated audio channels and polarity for data concepts~\cite{ferguson2017evaluation, hermann2011sonification, walker2002magnitude, walker2005mappings, wang2022seeing, batterman2012auditory, batterman2013auditory}. For example, loudness maps well to uncertainty~\cite{ballatore2019sonifying}. In our research, we extend their work by incorporating visual channels. 

% mental models
Finally, it is not well understood how users employ mental models when perceiving sonification \cite{kramer1994intro, walker2007consistency}. Since this is important for designing sonifications \cite{Ferguson2018congruence}, we provide an initial insight into strategies users employ when pairing audio with visual channels and mapping uncertainty to AV channels.

\subsection{Audiovisual Analytics}

% AV benefits
There are many examples that combine visualization and sonification in the AV analytics community \cite{lenzi2021data, Yoon2017multimodal}. However, there is little theory researchers can rely on \cite{enge2023towards, enge2024open}. While AV pairs, such as \embraced{y-position/pitch} and \embraced{visual density/harmonic density}, are commonly used, they might not be suitable for all data concepts~\cite{caiola2022audiovisual}, including uncertainty. Hence, our research contributes to the theoretical foundation regarding AV mapping choices for quantified~uncertainty. 

% benefits - non redundent
Empirical work has shown that combining sonification with visualization enhances data processing, anomaly detection, contrast perception, and target identification \cite{ronnberg2019sonification, cooke2017exploring, hildebrandt2016combining, rubab2023exploring}. Redundant AV mappings aid monitoring tasks, particularly in low-vision conditions, and enable vision-impaired users to collaborate with sighted individuals \cite{peres2005auditory, cooke2017exploring}. It also helps manage complex visualizations like parallel coordinates plots without overloading users \cite{ronnberg2016sonification, kramer2010sonification}. AV analytics is especially useful for scientific data, which often includes uncertainty \cite{enge2023towards}. For instance, Bearman \cite{bearman2011using} demonstrated that encoding uncertainty redundantly via auditory and visual channels improved efficiency and user preference compared to sonification~alone.

Rather than highlighting the benefits of AV representations, our research focuses on combining auditory and visual channels to communicate uncertainty.

%%%%%%%%%% Experimental setup %%%%%%%%%%%%%%%

\section{Experimental Setup}   
\label{setup}
We designed two crowd-sourced user studies: Experiment~1 to answer \textbf{RQ1} and Experiment~2 to answer \textbf{RQ2}. Both user studies were integrated into one experimental setup. The studies were preregistered on OSF.\footnote{\url{https://doi.org/10.17605/OSF.IO/VE7T8}} Participants first carried out Experiment~1, followed by Experiment~2. The stimuli and scripts used for the experiment can be found in the supplemental materials \cite{Vriend:2025:DarusSupplemental}.

\subsection{Stimuli}
% visual stimuli

We chose stimuli that can be created using code-based generative models. This approach allows precise control over the stimulus characteristics, ensuring consistency across experimental conditions and supporting reproducibility~\cite{schulz2016generative}. 

We focus on exploring AV pair polarities and the polarity of AV mappings of uncertainty. We examine whether users prefer an audio channel increase to correspond with a visual increase or decrease and how they map AV dimensions to probability. For each audio and visual channel, we generated three stimuli: an anchor, a high, and a low stimulus. During the experiments, these were presented as positive (anchor and high) or negative polarity (anchor and low).

\begin{figure}[h!]
\centering
\includegraphics[width=\columnwidth]{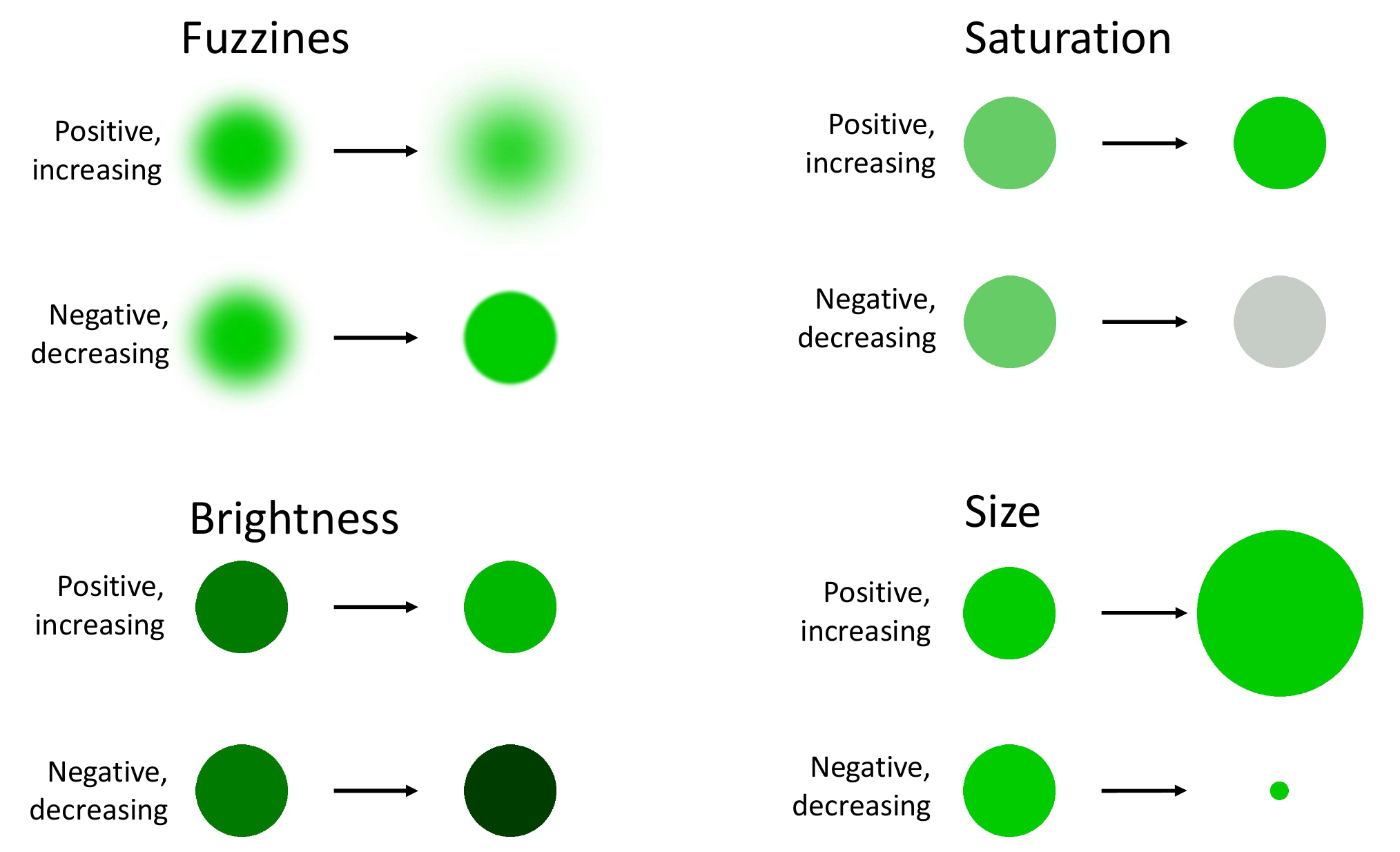}
\caption{The visual stimuli used in both experiments. The arrows indicate the polarities examined in the experiments for each channel. Recreated from MacEachren et al. \cite{maceachren2012visual}.}
\Description{This figure showcases four different visual properties—Fuzziness, Saturation, Brightness, and Size—demonstrating how they change from one state to another. Each property is represented by two circles connected by arrows, indicating the direction of change. Fuzziness (top left): Two green circles are displayed, with the one on the left being blurred, and the one on the right being sharp and well-defined. Saturation (top right): Two circles are shown transitioning in color. The left circle is a lighter green, and the right circle is more saturated, resulting in a deeper green. Another pair shows a lighter green transitioning to gray. Brightness (bottom left): Two green circles are shown, with the left circle being darker, and the right one brighter. Size (bottom right): The left circle is smaller, while the right circle is significantly larger, demonstrating a change in scale. Each section visually represents a different dimension of change, with arrows indicating the progression from one state to another.}
\label{fig:vis_stim}
\end{figure}

% visual
The visual channels include fuzziness, brightness, saturation, and size (see \Cref{fig:vis_stim}) as these are often used to visualize uncertainty \cite{kamal2021recent, padilla2020uncertainty} and intuitive mappings of uncertainty according to prior research~\cite{maceachren2012visual}. 

% sound stimuli
\begin{figure}[t]
\centering
\includegraphics[width=\columnwidth]{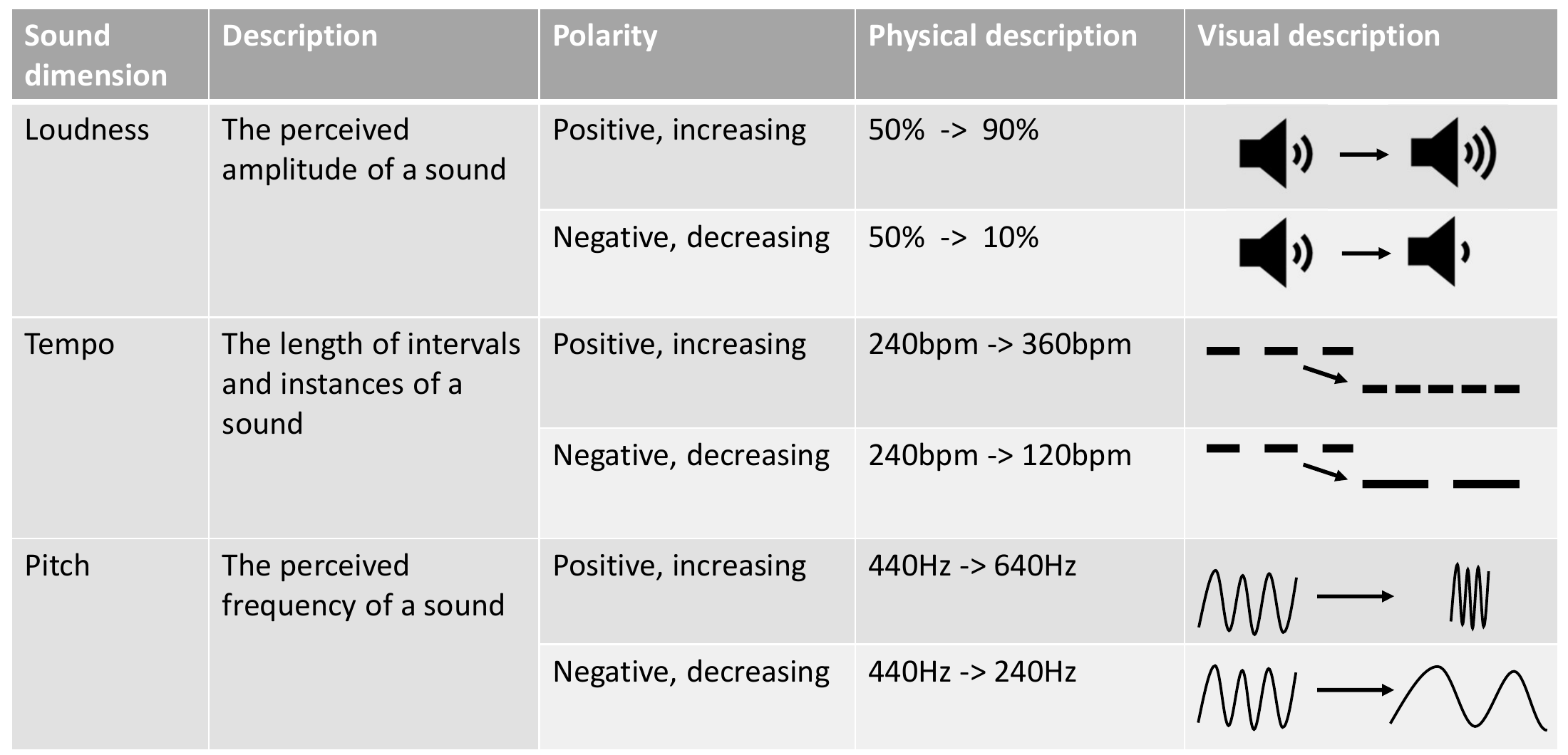}
\caption{A textual and visual description of the audio channels used in both experiments. The arrows under polarity and visual description indicate the polarities per channel. The figure was adapted from Ballatore et al.~\cite{ballatore2019sonifying}.}
\Description{This figure is a table that describes three sound channels: Loudness, tempo, and pitch. Each channel is presented with a description, a range of polarity, and a visual representation. Loudness: Description: The perceived amplitude of a sound. Polarity: Two examples are given: From 50\% to 90\%, From 50\% to 10\%. Visual Description: The visual representation shows two speaker icons in each example, with the one on the right emitting larger sound waves, indicating an increase in loudness, and the one on the bottom emitting smaller sound waves, showing a decrease in loudness. Tempo: Description: The length of a sound. Polarity: Two examples: From 240 beats per minute (bpm) to 360 bpm (increasing the tempo), from 240 bpm to 120 bpm (decreasing the tempo). Visual Description: The visual shows dashed lines representing tempos. In the top example, the tempo gets more frequent (shorter dashes), and in the bottom, the tempo slows down (longer spaces between dashes). Pitch: Description: The physical quality of the pitch of a sound. Polarity: Two examples: From 440 Hz to 640 Hz (increase in pitch), from 440 Hz to 240 Hz (decrease in pitch). Visual Description: The visual displays two piano keyboards in each example. In the top example, the second keyboard represents a higher pitch, and in the bottom example, the second keyboard represents a lower pitch. }
\label{fig:snd_stim}
\end{figure}
% sound
We chose pitch, tempo, and loudness for the audio channels because these are among the most used to sonify quantities~\cite{dubus2013systematic} and are considered suitable mappings of uncertainty~\cite{ballatore2019sonifying} (see \Cref{fig:snd_stim}). The auditory stimuli were 3-second-long sine waves to ensure the stimuli are encoded into echoic memory \cite{snyder2000music}. We generated the stimuli with equal physical distances between the levels of each dimension to ensure a perceivable difference. Since we cannot control the exact perceived quality of the loudness channel, we generated normalized relative loudness in SuperCollider\footnote{\url{https://supercollider.github.io/}} at 10\%, 50\%, and~90\%.

\subsection{Overview of Both Experiments}
\label{overview}

Experiment~1 and 2 employed an unspeeded variant of a speeded classification \cite{marks2004cross}. Speeded classification is used to understand cross-modal interactions between audio and visual stimuli \cite{gallace2006multisensory, Evans2010crossmodal}. Speeded classification is similar to an implicit association task \cite{greenwald1998measuring} as it asks participants to classify a stimulus rapidly. Participants perform faster when the AV dimensions match, indicating a cross-modal interaction \cite{marks2004cross}. Our unspeeded variant allowed more time per trial. Mappings with strong polarity agreement are likely more suitable due to reduced ambiguity in the represented data \cite{walker2002magnitude}.

Both experiments adopted a within-subjects design with a randomized trial order; all participants tested all AV stimuli combinations in random order. Participants received a short training session before each experiment, with separate stimuli in order to not influence the results. 

% recoding of mapping preference
% We were interested in congruency preference: how the polarities of the channels should match.
We reverse-coded the responses to match the positive audio polarity. For example, a participant prefers \embraced{loudness\polardown \polarup fuzzy}. After reverse-coding, this becomes a preference for \embraced{loudness\polarup \polardown fuzzy}. In this example, \embraced{loudness\polarup \polardown fuzzy} is preferred when the polarities do not match. Reverse-coding relies on a participant having a consistent preference regarding an AV pair or mapping. For example, if a participant prefers \embraced{brightness\polardown \polardown pitch}, they should also prefer \embraced{brightness\polarup \polarup pitch}; otherwise, their preference would not be consistent. Reverse-coding is common practice in multi-sensory \cite{FREEMAN202066} and human-computer interaction research \cite{Anwar2023haptic} to, for example, deal with negatively phrased questions. We reverse-coded the stimuli to a positive polarity to simplify the analysis and make the results easier to read. 

% measurements and analysis
% We analyzed the participants' preferences, RT, and mapping strategies in both user studies. 
% mapping preference 
Following \citet{ballatore2019sonifying}, we used an exact binomial test to analyze polarity preferences for each AV stimulus. With this test, we measure participant agreement on polarity mappings. For example, if 90 out of 100 participants paired increasing size with increasing loudness, the binomial probability is 0.9, indicating a strong preference for \embraced{loudness\polarup \polarup size}. In contrast, a 50\% split (e.g., size with pitch) yields a probability of 0.5, suggesting \embraced{pitch/size} is not a strong AV pair.

% reaction times (RT)
We recorded the reaction times (RTs) in milliseconds that participants needed to indicate their preferences for a given AV pair or mapping. We calculated the mean RT per participant per combination of conditions. 

% mapping strategies
After participants completed each experiment, we asked them to rate the task's difficulty on a 7-point Likert scale and to report their mapping strategy. These results were analyzed qualitatively.

\subsection{Participants}
Through Prolific, 217 participants participated online, using their own PCs and headphones. All participants were at least 18 years old, fluent in English, reported to have perfect or corrected vision, and no hearing difficulties. They were paid £9 per hour, according to Prolific's guidelines, regardless of the quality of their data.

The participants had diverse nationalities, genders, ages, and employment statuses. Their average age was 33 years, ranging from 18 to 76 years. 34.6\% identified as female and 65.4\% as male. English was the mother tongue of 45.9\% of the participants, while 54.1\% spoke English as a second language. The participants held nationalities such as British, Polish, Portuguese, South African, or~others.

%%%%%% Experiment 1 content %%%%%%%%%%%%%%%

\section{Experiment~1: Audio/Visual Pairs}

\begin{figure}[t]
\centering
\includegraphics[width=\columnwidth]{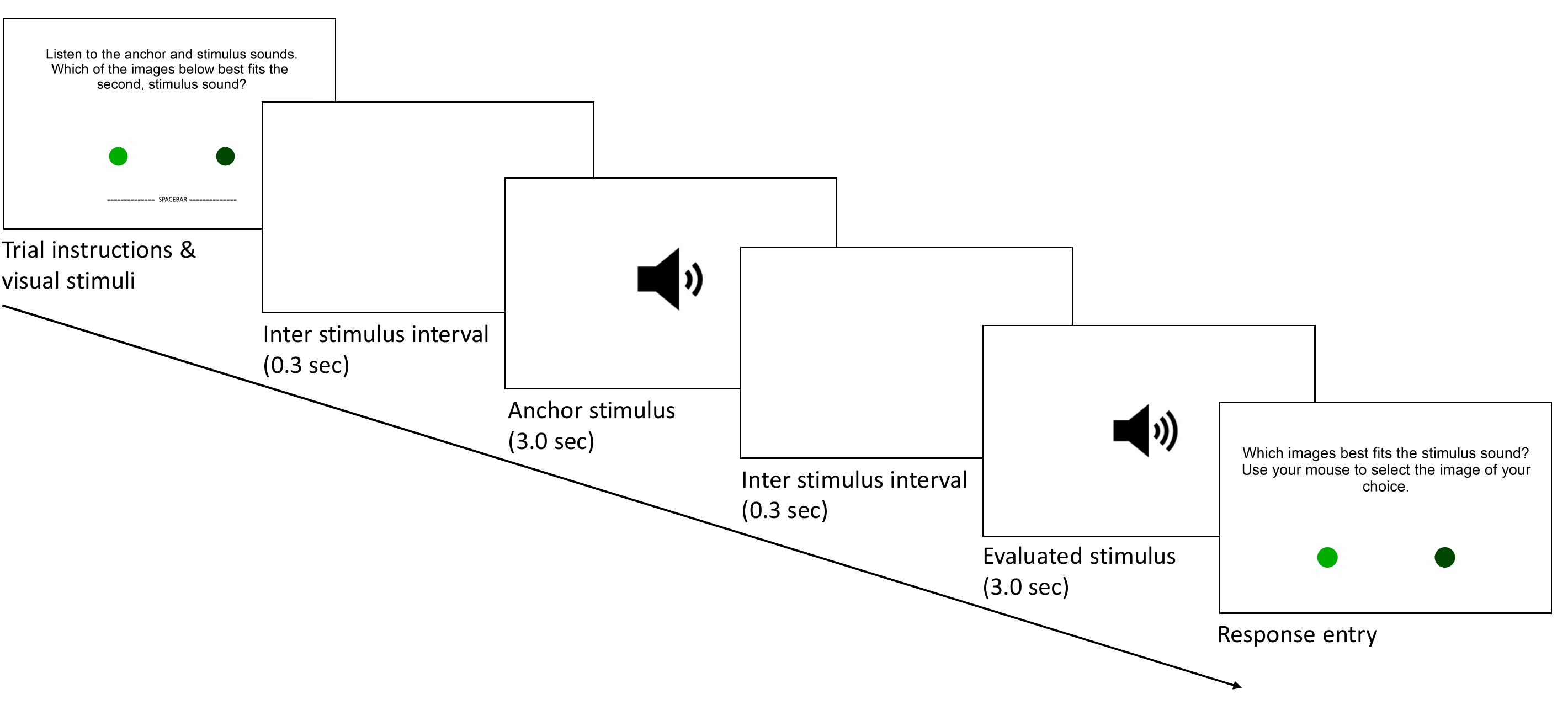}
\caption{The order of trials in Experiment~1. Participants were first shown two visual stimuli. The visual stimuli were variants of the same channel (e.g., big-sized and small-sized circles). Once ready, participants pressed a button to continue, which removed the visual stimuli. Participants then listened to two audio stimuli separated by a 0.3-second interval. The audio stimuli were variants of the same channel (e.g., high pitch and low pitch). The visuals were then displayed again, and participants selected which visual better represented the second sound they heard.}
\Description{This figure shows a diagram of the trial order of this experiment. The diagram progresses diagonally from the top left toward the bottom right corner with an arrow indicating the progression of the trial order. The diagram starts in the top left corner with a screenshot of "Trial instructions \& visual stimuli". The screenshot shows unreadable text and examples of two visual stimuli side by side. Next, it shows a blank screenshot with the label "inter stimulus interval (0.3 sec)". The next screenshot shows an icon of a speaker and the label reads "First audio stimulus (0.3 sec)". Next, it shows a blank screenshot with the label "inter stimulus interval (0.3 sec)". Next, it shows a blank screenshot with the label "inter stimulus interval (0.3 sec)". The next screenshot shows an icon of a speaker, and the label reads "Second audio stimulus (0.3 sec)". The next screenshot with the label "response entry" shows unreadable text with two visual stimuli side by side. The last screenshot shows a blank screenshot with the label "inter stimulus interval (0.3 sec)". }
\label{fig:task1}
\end{figure}

Experiment~1 investigated \textbf{RQ1: Which pairs of audio and visual stimuli are most preferred?} using a within-subject design with three factors: sound dimension, sound polarity, and visual dimension. \textbf{The visual dimension has four levels (fuzziness, brightness, size, saturation; see \Cref{fig:vis_stim}).} Similarly, \textbf{the auditory dimension has three levels (loudness, tempo, pitch; see \Cref{fig:snd_stim}), and sound polarity has two levels (increasing or decreasing).} We randomized the factors. The trials randomized the three factors per participant. The trials were presented to the participants as explained in \Cref{fig:task1}. The participant's task was to evaluate the second audio stimulus in relation to the first and to choose the visual stimulus that best suited the second stimulus. Participants completed 48 trials. 

We measured the polarity preference per AV pair. With polarity preference, we mean whether participants prefer an increase in the visual channel paired with an increase (positive polarity) or a decrease (negative polarity) in the audio channel. The more participants agree on the polarity preference of a visual and audio channel, the higher the polarity preference of that pairing. On the contrary, there is no agreement if half of the participants prefer a positive polarity, while the other half a negative polarity for a given AV pair. Preferred AV pairs are pairs that lead to a high agreement of polarity preference. Since an agreement in many AV cross-modal correspondences exists \cite{parise2013audiovisual}, there will be a stronger polarity preference for some AV pairs compared to others. We used an exact binomial test to test the extent of polarity preference for each AV pair against random preference. After reverse-coding, there are two polarity preferences for each AV pair: both visual and audio channels increase, or the visual channel increases while the audio channel decreases. Hence, random preference means 50\%, or a 0.5 proportion. This leads to the following hypothesis: \textbf{Per AV pair, we hypothesize that the proportion of the polarity preferences significantly differs from 0.5.}

Additionally, we examined the RT per AV pair. We expect the audio and visual channels to affect RT. Thus, our hypothesis is: \textbf{There is an interaction between audio and visual channels for RT.}

\subsection{Results}
% prep data
We excluded participants with outlier data and those who did not pass the audio and visual comparison check. The sample size of Experiment~1 was 170.

\subsubsection{Audio/visual Pair Preferences}
% analysis
% results
\begin{figure}[h!]
\centering
\includegraphics[width=\columnwidth]{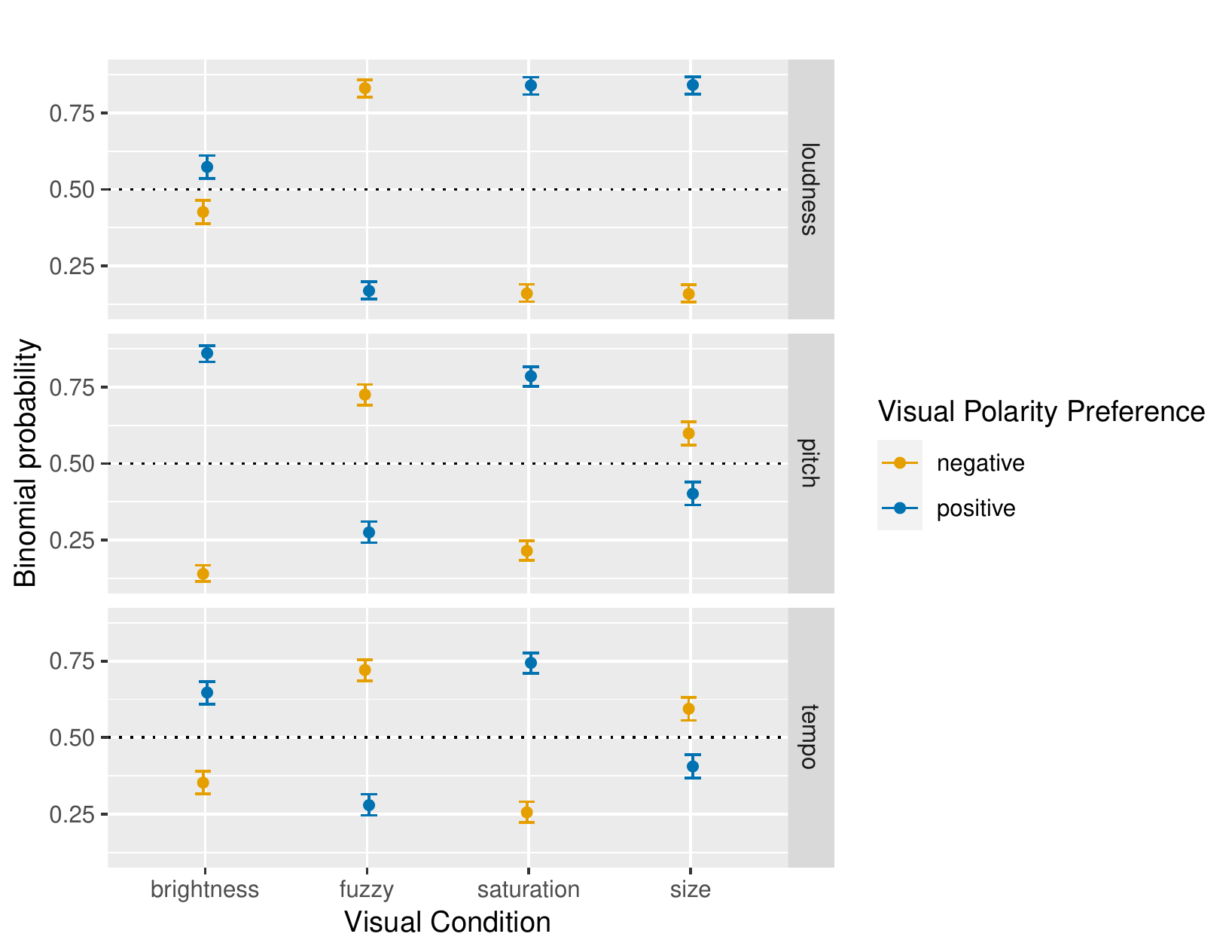}
\caption{The mean proportion per visual and audio condition for each polarity preference. The error bars show the 95\% confidence intervals. The further away from the reference line (binomial probability of 0.5), the stronger the preference is for that AV pair. }
\Description{This image shows the 95\% confidence intervals of the exact binomial tests. The graph is divided, vertically, into three sub-graphs, each representing an audio condition. From top to bottom: pitch, loudness, tempo. The x-axis represents the visual conditions, which are brightness, fuzzy, saturation, and size. The y-axis represents the binomial probability, ranging from 0 to 0.75. Each error bar in the graph represents the 95\% confidence interval of the binomial probability for a given visual and audio condition. The bars are colored blue and orange, with blue bars indicating positive visual polarity preference and orange bars indicating negative visual polarity preference. }
\label{fig:exp1_binom}
\end{figure}

% results
We conducted an exact binomial test per AV pair to test the hypothesis: \textbf{The proportion of the polarity preferences significantly differs from 0.5.} The proportion of all visual polarity preferences is significantly different from 0.5, hence we reject $H0$ (full results are in Appendix~\ref{exp1_mapping}). According to \Cref{fig:exp1_binom}, some pairs showed a stronger polarity preference than others. The participants preferred \embraced{pitch\polarup \polarup brightness} (binom prop. of 0.86), \embraced{loudness\polarup \polarup saturation} (binom prop. of 0.84), \embraced{loudness\polarup \polarup size} (binom prop. of 0.84), and \embraced{loudness\polarup \polardown fuzziness} (binom prop. of 0.83).

\subsubsection{Reaction Time}
% prep the data
% We aggregated the data so we had one value per participant per audio condition, per visual condition. 
According to the Shapiro-Wilk test, the RT data deviated significantly from a normal distribution for all AV pairs ($p < 0.001$). We applied a log transformation, after which the data per AV pair was found to be normally distributed.

% descriptives
\begin{figure}[h!]
\centering
\includegraphics[width=\columnwidth]{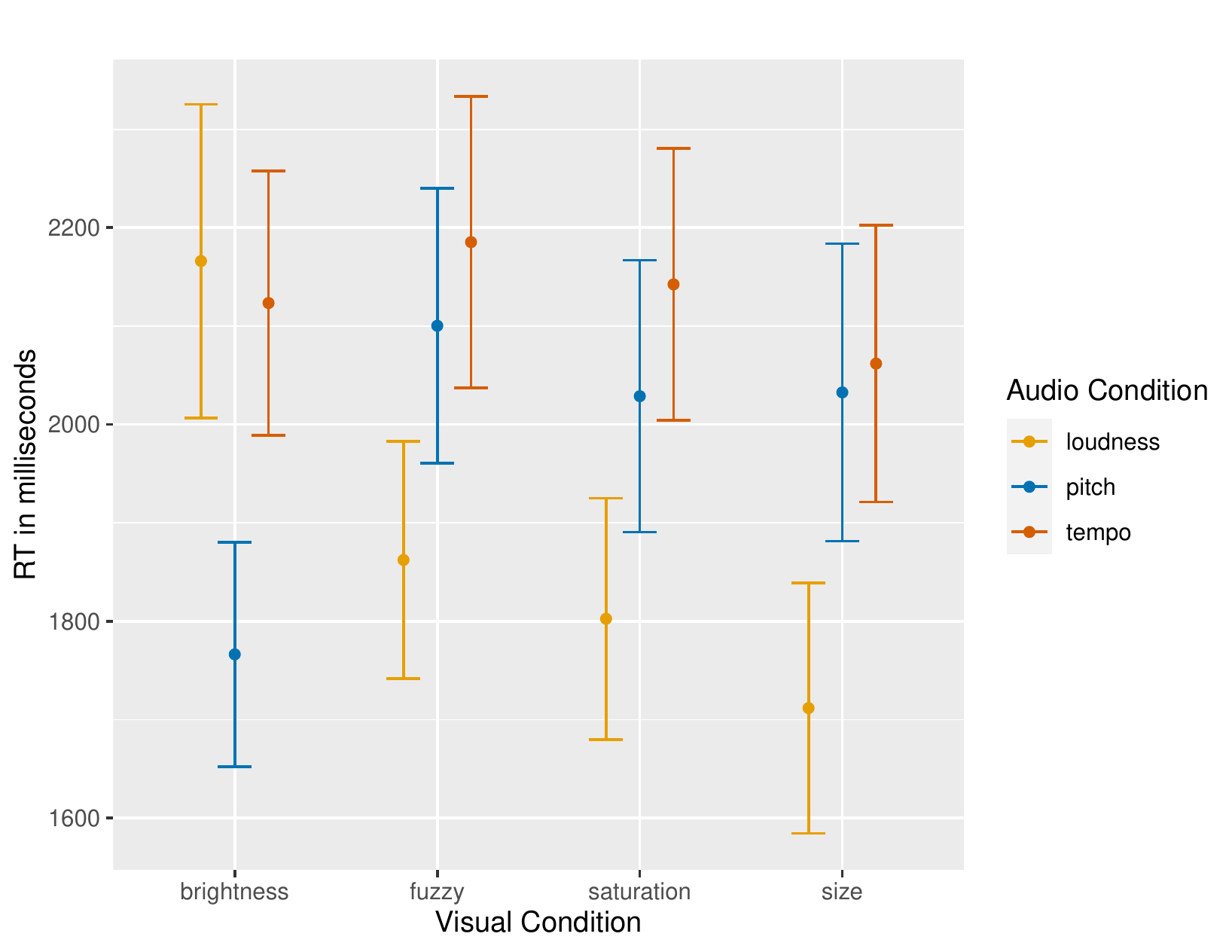}
\caption{The mean RT in milliseconds per visual and audio channel. The error bars represent the 95\% CI. The lower the RT, the quicker the participants were to indicate their preference.
}
\Description{This image shows 95\% confidence intervals of the effects of visual and audio conditions on reaction time (RT). The x-axis represents the visual conditions, which are brightness, fuzzy, saturation, and size. The y-axis represents the reaction time in milliseconds, ranging from 1600 to 2200. Each bar in the graph represents the 95\% CI of the average reaction time for a given visual and audio condition. The bars are grouped by color: blue, orange, and red, with blue bars representing the pitch audio condition, orange bars representing the loudness audio condition, and red bars representing the tempo audio condition.}
\label{fig:exp1_CI}
\end{figure}

\begin{figure}[h!]
\centering
\includegraphics[width=\columnwidth]{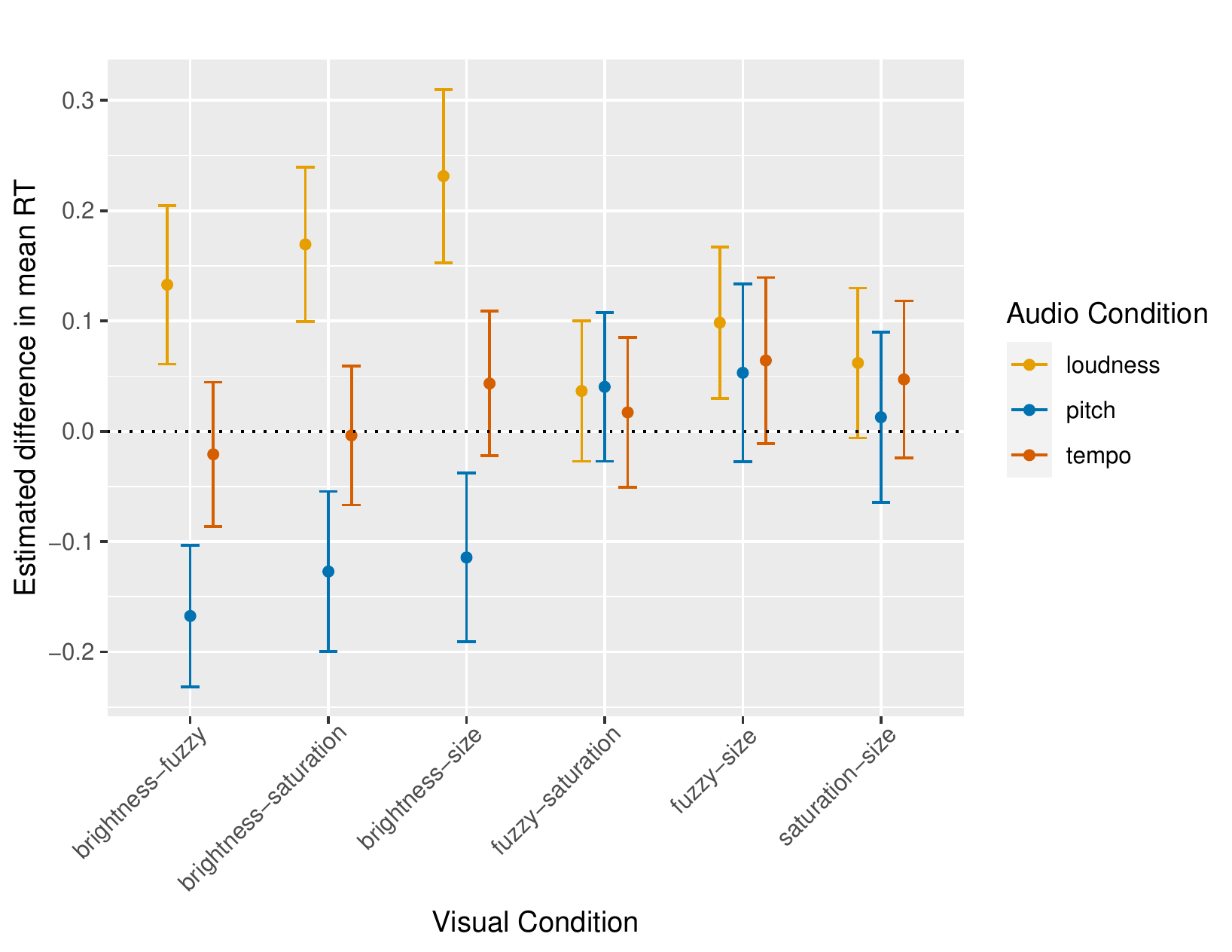}
\caption{The estimated difference in transformed mean RT of each AV pair, tested between each visual dimension, at the audio dimension level. The error bars show the 95\% CI. The further away from the reference line (0) the CI is, the bigger the difference in RT between the pairs. }
\Description{This image shows a graph of the 95\% confidence intervals of the pairwise analysis of the average reaction times per audio and visual stimulus. The comparison was done on the level of the visual conditions written on the bottom x-axis: from left to right: brightness-fuzzy, brightness-saturation, brightness-size, fuzzy-saturation, fuzzy-size, saturation-size. The bars are divided into three categories, by color, that represent different audio conditions: pitch, loudness, and tempo. If the confidence intervals overlap between conditions, that suggests that they could be due to chance, and there may not be a real difference between the conditions.}
\label{fig:exp1_pair}
\end{figure}
  
% analysis
We performed a two-way within-subjects ANOVA (3 $\times$ 4) to test the hypothesis: \textbf{There is an interaction between audio and visual channels for RT}. There was a statistically significant interaction between the effects of the sound channel and the visual channel ($F(6, 1014)=10.26, p<0.001, \eta_{p}^{2}=0.057$).

% simple main effects
A simple main effect with Bonferroni correction shows a significant result of the visual condition on pitch ($F(2.84, 480)=7.47, p=0.006, \eta_{p}^{2}=0.042$) and loudness ($F(3, 507)=15.10, p<0.001, \eta_{p}^{2} = 0.082$). A Bonferroni-corrected simple main effect showed significant effects of the audio condition on brightness ($F(2, 338)=18.60, p<0.001, \eta_{p}^{2}=0.099$), fuzziness ($F(2, 338)=10.30, p=0.004, \eta_{p}^{2}=0.057$), saturation ($F(2, 338)=12.30, p<0.001, \eta_{p}^{2}=0.068$), and size ($F(2, 338)=12.30, p<0.001, \eta_{p}^{2}=0.068$).

% pairwise analysis
\Cref{fig:exp1_CI} shows that \embraced{loudness/size} and \embraced{pitch/brightness} resulted in the lowest RT. A Bonferroni-corrected pairwise analysis shows a significant difference in the pitch and loudness pairs. As seen in \Cref{fig:exp1_pair}, \embraced{pitch/brightness} leads to a lower RT compared to \embraced{pitch/fuzziness} ($p<0.001$), \embraced{pitch/saturation} ($p=0.005$), and \embraced{pitch/size} ($p=0.02$). \embraced{Loudness/brightness} leads to a significantly higher RT compared to \embraced{loudness/fuzziness} ($p=0.003$), \embraced{loudness/saturation} ($p<0.001$), and \embraced{loudness/size} ($p<0.001$). The full results can be found in Appendix~\ref{exp1_rt_pair}.

\subsubsection{Mapping Strategies}
% task difficulty

\begin{figure}[t]
\centering
\includegraphics[width=0.8\columnwidth]{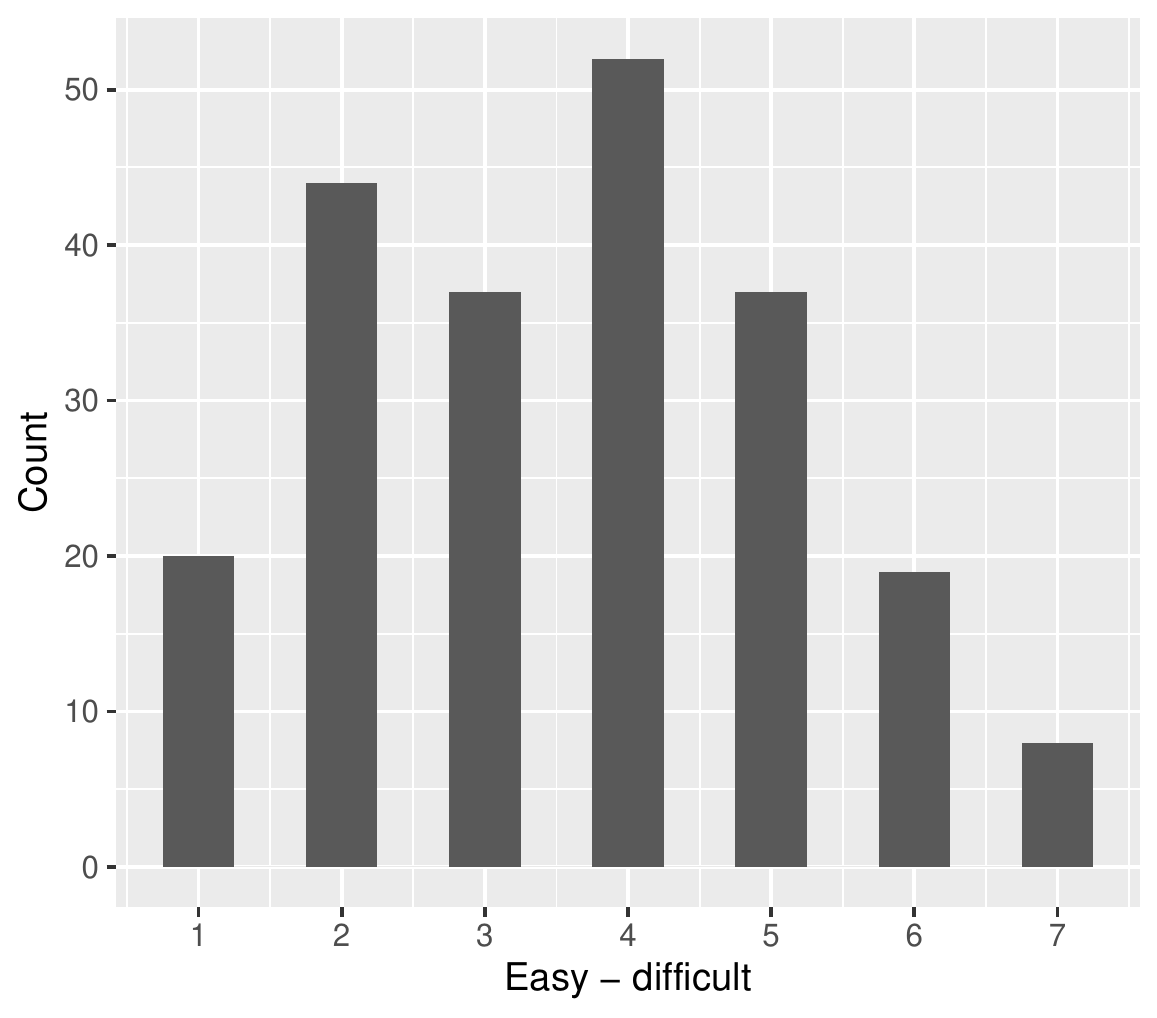}
\caption{A histogram of the difficulty of the task in Experiment~1, according to ratings by participants, on a 7-point Likert scale (from 1 = ``easy''  to 7 =  ``difficult''). The shape of the histogram is slightly skewed to the left, indicating that participants thought the task was easy to moderately difficult.}
\Description{The image shows a histogram of the distribution of responses on a scale from 1 to 7, representing levels of difficulty from easy to difficult. The x-axis represents the scale values: 1 to 7 (easy to difficult), while the y-axis represents the count of responses for each value. The overall shape of the histogram is skewed to the right, indicating that most responses are on the easier end of the scale. There is a clear peak at the value of 4, suggesting that this is the most common response. The histogram also shows that there are fewer responses at the higher end of the scale, indicating that the task was generally perceived as easy to moderate in difficulty.}
\label{fig:exp1_diff}
\end{figure}

% difficulty
\Cref{fig:exp1_diff} shows that participants mostly thought the task was easy to intermediately difficult.

% audio-visual
According to the free-text answers, most participants \textbf{focused on both audio and visual attributes} (60) or \textbf{focused primarily on sound attributes} (50). 

One participant who \textbf{focused on both audio and visual attributes} used mental imagery to map the audio to the visual: \textit{``I associated the sound with the circle [by] making a mental image. For example, the faded circles would be slower-paced and quieter, whereas a brighter green would be more energetic and faster and louder. If a circle was larger I would associate that with more noise than a smaller one.''} Another participant focused on the more qualitative and subjective attributes of the stimuli: \textit{``I paid attention to associations. The more pleasant sounds had a dark green color, which I prefer. The more delicate, muffled sounds had a blurry surface. [...].''} Of those participants who \textbf{focused on both audio and visual attributes} we noticed that six participants were looking for \textbf{common features} between the audio and visual stimuli: \textit{``[I] tried to find what comparison I could make between the two [sounds]. [...] Then compared the shapes similarly, visually in my mind.''}

% audio only
Some participants who \textbf{focused primarily on audio attributes} also focused more on the qualitative and subjective sides of the audio stimuli: \textit{``I assessed whether the sound was stronger or weaker.''} Others focused on the differences between the two stimuli presented in each trial: \textit{``[I] tried to listen to the difference between the two sounds and then tried to imagine how I would represent the sounds visually.''} We noticed that three participants who \textbf{focused primarily on audio attributes} mentioned that they listened to the intensity of the audio stimuli: \textit{``I related the higher intensity pitch sounds to the larger sizes brighter color circles.''}

% visual only 
18 participants \textbf{focused primarily on visual attributes}. One participant wrote: \textit{``I tried to imagine what these figures would sound like and choose based on that.''} Some intuition also came into play for one participant: \textit{``I just went with whatever felt more like the image. Deeper tones were usually larger images or deeper colors.''}

% something else
Five other participants did not map the audio stimuli directly to the visual stimuli but \textbf{mapped stimuli to something else first}. For instance, a participant wrote: \textit{``I tried to find a descriptive word for the contrast in sound and then find the most appropriate image for the word.''}

% scenarios
Yet, four participants mentioned that they made \textbf{use of scenarios}. One participant, for example, imagined the following: \textit{``I imagined [a] pulsing object on a radar and linked it with the sounds that I was hearing.''}

% different strat
Six participants varied their mapping strategy throughout the experiment: \textit{``I tried to vary [...] my strategies to find a [pattern]. [...] When the sound [was] continuous I would choose the biggest, the brightest, or most clear image. Other times I would choose at random.''}

% intuition
We found that 25 participants mentioned that they relied on \textbf{intuition}: \textit{``After hearing the sound samples I [selected] the image based [on] my gut feeling.''}

% unsure
We were \textbf{unsure} what kind of strategy four of the participants used. This could be because what they wrote did not relate entirely to the question. For example: \textit{``I felt like the pulsing sounds were easier to identify than the monotone sounds. [...] The blurred shapes were easier than the non-blurred, but no strategy as such.''}

\subsection{Discussion}
We found AV pairs with strong polarity preference agreement, indicating cross-modal correspondence \cite{parise2013audiovisual}. The RT analysis showed the same trend.

% preference
According to \Cref{fig:exp1_binom}, all pairs showed a significant polarity preference agreement. However, we noticed a particularly strong polarity preference for \embraced{pitch\polarup \polarup brightness}, followed by \embraced{loudness\polarup \polarup size} and \embraced{loudness\polarup \polarup saturation}. These pairs are preferred when visual and audio dimensions increase (or decrease). Another preferred pair was \embraced{loudness\polarup \polardown fuzziness}. Here, an increase in loudness is best paired with a decrease in fuzziness, or vice versa. 

% RT
According to \Cref{fig:exp1_CI}, preferred pairs also lead to faster RT. This aligns with the speeded classification paradigm, which states that participants will typically perform faster when the AV dimensions match \cite{marks2004cross}. While the \embraced{loudness/size} pair led to the lowest mean RT, we only found a significant difference between size and brightness for the loudness pairs. However, this makes sense since \embraced{loudness/saturation} and \embraced{loudness/fuzziness} also led to a low RT. The \embraced{pitch/brightness} pair led to a low RT. This was a significantly faster pair for loudness compared to saturation, fuzziness, and size.

% mapping
According to the analysis of the mapping strategies, many participants could process the AV pairs holistically. However, most participants reported evaluating the audio and visual stimuli separately by either focusing on the audio or visual stimuli.

% general
The following AV pairs led to the highest polarity agreement and the lowest RT: \embraced{loudness\polarup \polarup size}, \embraced{loudness\polarup \polarup saturation}, 
\inlinemaths{\langle}pitch\polarup \mbox{} \polarup  brightness\inlinemaths{\rangle},
and \embraced{loudness\polarup \polardown fuzziness}.

%%%%%%%%% Experiment 2 content %%%%%%%%%%%%%%%%%%%%%

\section{Experiment~2: Uncertainty Mappings} 
\label{exp2_intro}

Experiment~2 explored the AV pairs from Experiment~1 as mappings of uncertainty with \textbf{RQ2: Which audiovisual mappings are most preferred to represent uncertainty?} The experimental design largely follows Experiment~1. However, instead of pairing audio and visual stimuli, participants were asked to map AV stimuli to probability. For the audio and visual dimensions, participants separately chose between ``high probability'' and ``low probability.'' Before the trials began, participants received the following explanation: \textit{Probability is a way to measure how likely something is to happen. It is often represented by percentages between 0\% and 100\%, where 0\% indicates that an event that is unlikely to happen and 100\% indicates that is certain to happen. For example, the weather forecast predicts a 30\% chance of rain tomorrow. This means that there is a 3 out of 10 chance that it will rain and there is a 7 out of 10 chance it will not rain. Thus, there is a low probability that it will rain tomorrow.} To ensure all participants understood ``probability'', we asked a multiple-choice question.

\begin{figure}[t]
\centering
\includegraphics[width=\columnwidth]{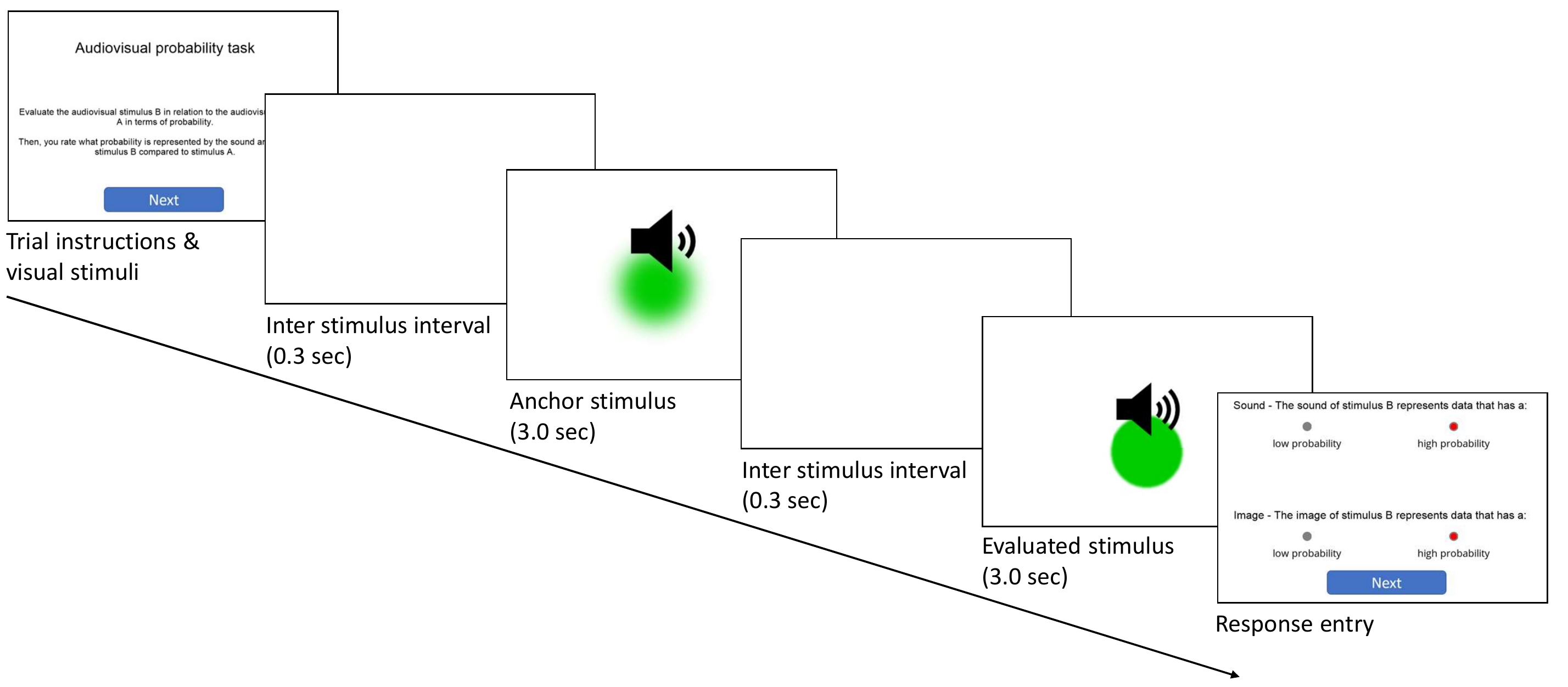}
\caption{The trial order in Experiment~2. Participants were first shown instructions. Once ready, participants pressed a button to continue, which removed the instructions. Participants were then presented with two AV stimuli separated by a 0.3-second interval. The AV stimuli were of the same dimensions but varied in polarity. Then the responses were collected; participants selected whether the second AV stimulus represented a high or low probability.}
\Description{This figure shows a diagram of the trial order of this experiment. The diagram progresses diagonally from the top left toward the bottom right corner with an arrow indicating the progression of the trial order. The diagram starts in the top left corner with a screenshot of "Trial instructions". The screenshot shows unreadable text. Next, it shows a blank screenshot with the label "inter stimulus interval (0.3 sec)". The next screenshot shows an icon of a speaker that overlaps an example of a visual stimulus, and the label reads "First audiovisual stimulus (0.3 sec)". Next, it shows a blank screenshot with the label "inter stimulus interval (0.3 sec)". Next, it shows a blank screenshot with the label "inter stimulus interval (0.3 sec)". The next screenshot shows an icon of a speaker that overlaps an example of a visual stimulus, and the label reads "Second audio stimulus (0.3 sec)". The next screenshot with the label "response entry" shows unreadable text with four radio buttons. The last screenshot shows a blank screenshot with the label "inter stimulus interval (0.3 sec)". }
\label{fig:task2}
\end{figure}

The experiment design was within-subject with four factors: audio dimension, audio polarity, visual dimension, and visual polarity. The audio dimension has three levels (pitch, loudness, and tempo; see \Cref{fig:snd_stim}), the audio polarity has two levels (decreasing and increasing), the visual dimension has four levels (fuzziness, brightness, saturation, and size; see \Cref{fig:vis_stim}), and the visual polarity has two levels (increasing and decreasing). We measured both the polarity preference and RT. The trials randomized the four factors per participant. The trials were presented as shown in \Cref{fig:task2}. Each participant completed 48 trials.

Polarity preference was assessed according to the hypothesis: \textbf{Per AV stimulus channel, we hypothesize that the proportion of the polarity preferences significantly differs from 0.25.} This hypothesis is based on the fact that there are four possible polarity preferences for each AV mapping. If there is no preference agreement, each polarity is preferred by 25\% of the participants. For example, participants could prefer to map increasing probability to \embraced{loudness\polarup \polarup fuzziness},  \embraced{loudness\polarup \polardown fuzziness}, \embraced{loudness\polardown \polarup fuzziness}, or \embraced{loudness\polardown \polardown fuzziness}.

The RT was examined for each mapping with the hypothesis: \textbf{There is an interaction between audio channel, visual channel, and sound/visual polarity mapping on reaction time~(RT).}

\subsection{Results} \label{exp2_analysis}
% data prep
% removed participants
We excluded participants according to the same criteria as Experiment~1. Additionally, we removed participants who answered wrongly twice to the probability multiple-choice question, since these participants did not demonstrate an understanding of probability. The sample size was 185.

\subsubsection{AV Mapping Preferences}
% Descriptives
\begin{figure}[t]
\centering
\includegraphics[width=\columnwidth]{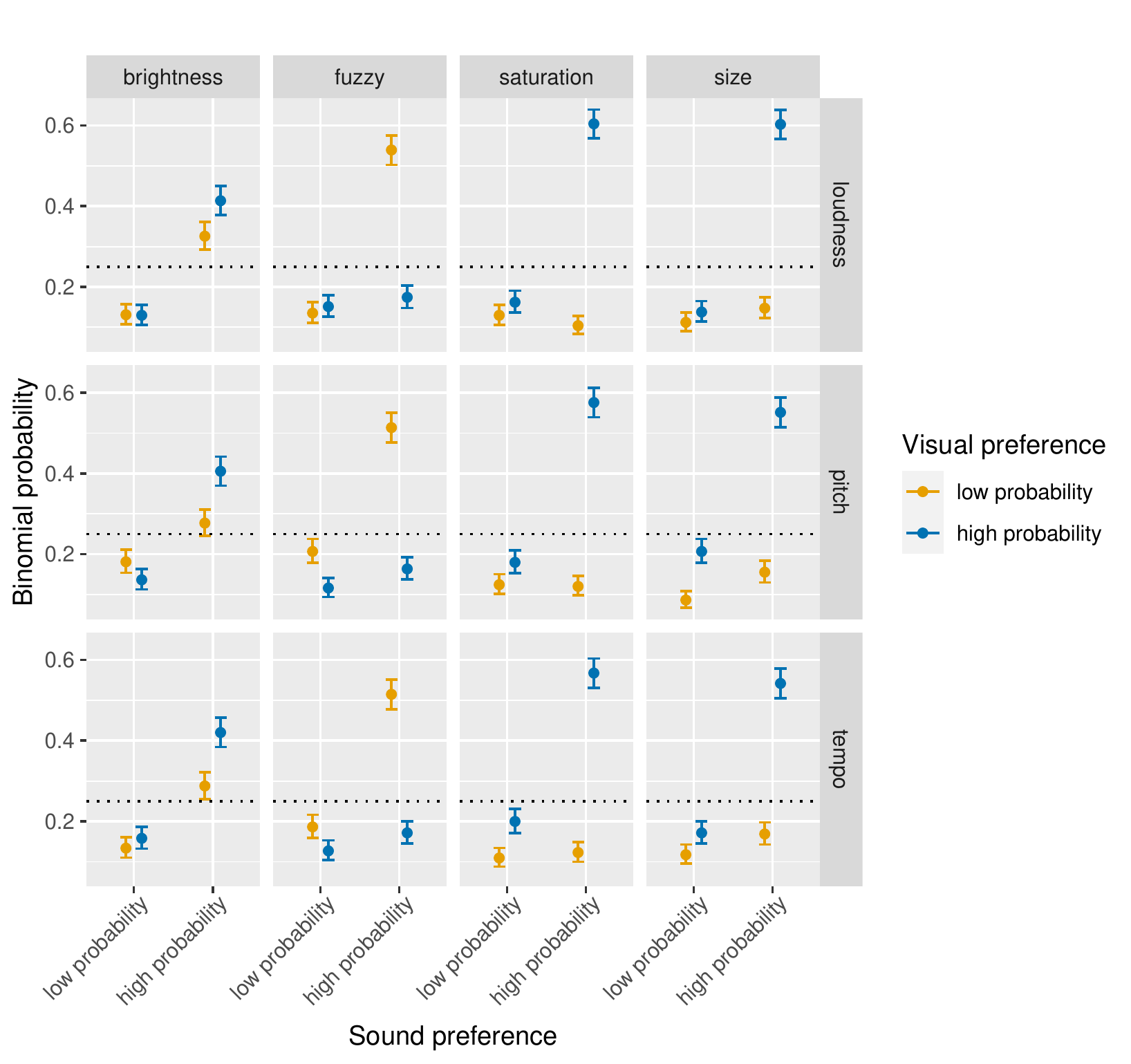}
\caption{The mean proportion for each probability polarity mapping preference per AV mapping. The error bars show 95\% confidence intervals. The further away from the reference line (binomial probability of 0.25), the stronger the preference is for that AV mapping.
 }
\Description{The image shows a series of 95\% CI showing the relationship between visual dimensions, audio dimensions, and binomial probability of polarity mapping preference. The graph is divided into sub-graphs. The audio dimension is stacked vertically and reads from top to bottom: pitch, loudness, tempo. The visual condition is stacked horizontally and read from left to right: brightness, fuzzy, saturation, size. The y-axis represents the binomial probability, ranging from 0 to 0.6. Each 95\% CI in the plot represents the binomial probability for a given combination of visual condition and sound preference. The dots are colored blue and orange, with blue dots indicating preference for high probability sound and orange dots indicating preference for low probability sound. }
\label{fig:exp2_binom}
\end{figure}

% analysis
We conducted an exact binomial test per AV pair to test the hypothesis: 
\textbf{The proportion of the polarity preferences differs from 0.25.} The proportions of all AV-probability mapping preferences are significantly different from 0.25, except for \embraced{pitch\polarup \polardown brightness}. See Appendix~\ref{exp2_mapping} for the results of the exact binomial tests. 

\Cref{fig:exp2_binom} shows that some mappings resulted in a strong preference. These included \embraced{loudness\polarup \polarup saturation} (binom prop. 0.60), \embraced{loudness\polarup \polarup size} (binom prop. 0.60), 
\embraced{pitch\polarup \polarup saturation} (binom prop. 0.58),
\embraced{tempo\polarup \polarup saturation} (binom prop. 0.57), and 
\inlinemaths{\langle}pitch\polarup \mbox{} \polarup  size\inlinemaths{\rangle}
(binom prop. 0.55). 

%\inlinemaths{\langle}pitch\polarup \mbox{} \polarup  brightness\inlinemaths{\rangle},

% Sorted top entries from the table in the appendix:
% loudness positive saturation positive 0.604
% loudness positive size positive 0.603
% frequency positive saturation positive 0.576 
% tempo positive saturation positive 0.568
% frequency positive size positive 0.551
% tempo positive size positive 0.542 
% loudness positive fuzzy negative 0.539 
% tempo positive fuzzy negative 0.515
 %frequency positive fuzzy negative 0.514

\subsubsection{Reaction Time}

\begin{figure}[h!]
\centering
\includegraphics[width=\columnwidth]{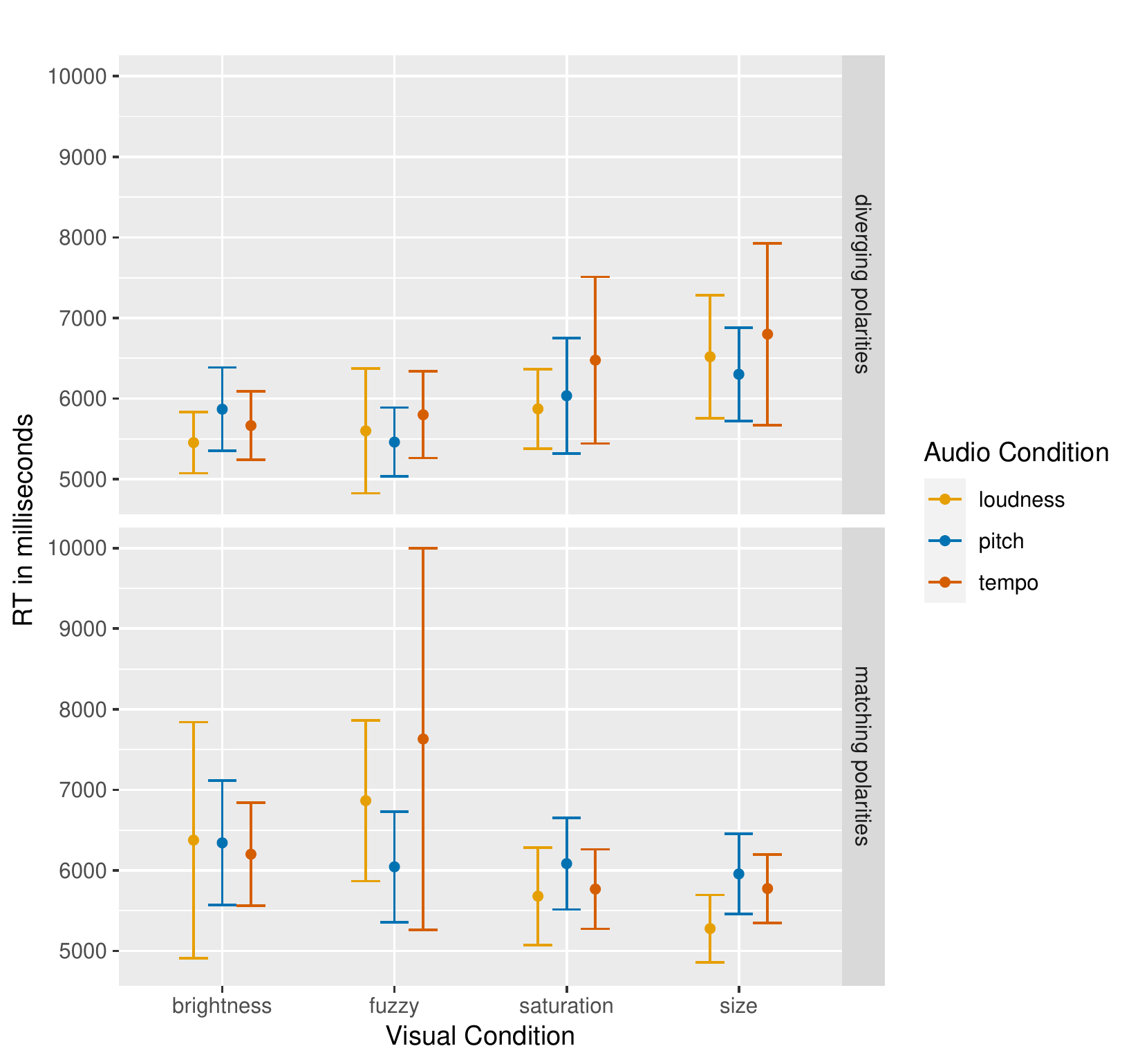}
\caption{The mean RT in milliseconds per visual and audio condition and per polarity mapping. The error bars represent the 95\% CI. The lower the RT, the quicker the participants were to indicate their preference.}
\Description{The image shows a series of 95\% CI graphs showing the reaction time results for each visual and audio condition per polarity mapping. The x-axis represents the visual conditions, which are brightness, fuzzy, saturation, and size. The y-axis represents the reaction time in milliseconds, ranging from 5000 to 10000. Each bar in the graph represents the average reaction time for a given combination of visual condition and audio condition. The bars are colored blue, orange, and red, with blue bars representing the pitch audio condition, orange bars representing the loudness audio condition, and red bars representing the tempo audio condition.  The graphs are organized into two rows, one for the matching polarities condition and one for the diverging polarities condition. The overall trend of the graphs is that reaction time tends to be higher for diverging polarities trials than for matching polarities trials, especially for the brightness, saturation, and size visual conditions. Additionally, the audio condition appears to affect reaction time, with some combinations of visual conditions and audio conditions leading to faster or slower reaction times.}
\label{fig:exp2_ci_raw}
\end{figure}

\begin{figure}[h!]
\centering
\includegraphics[width=\columnwidth]{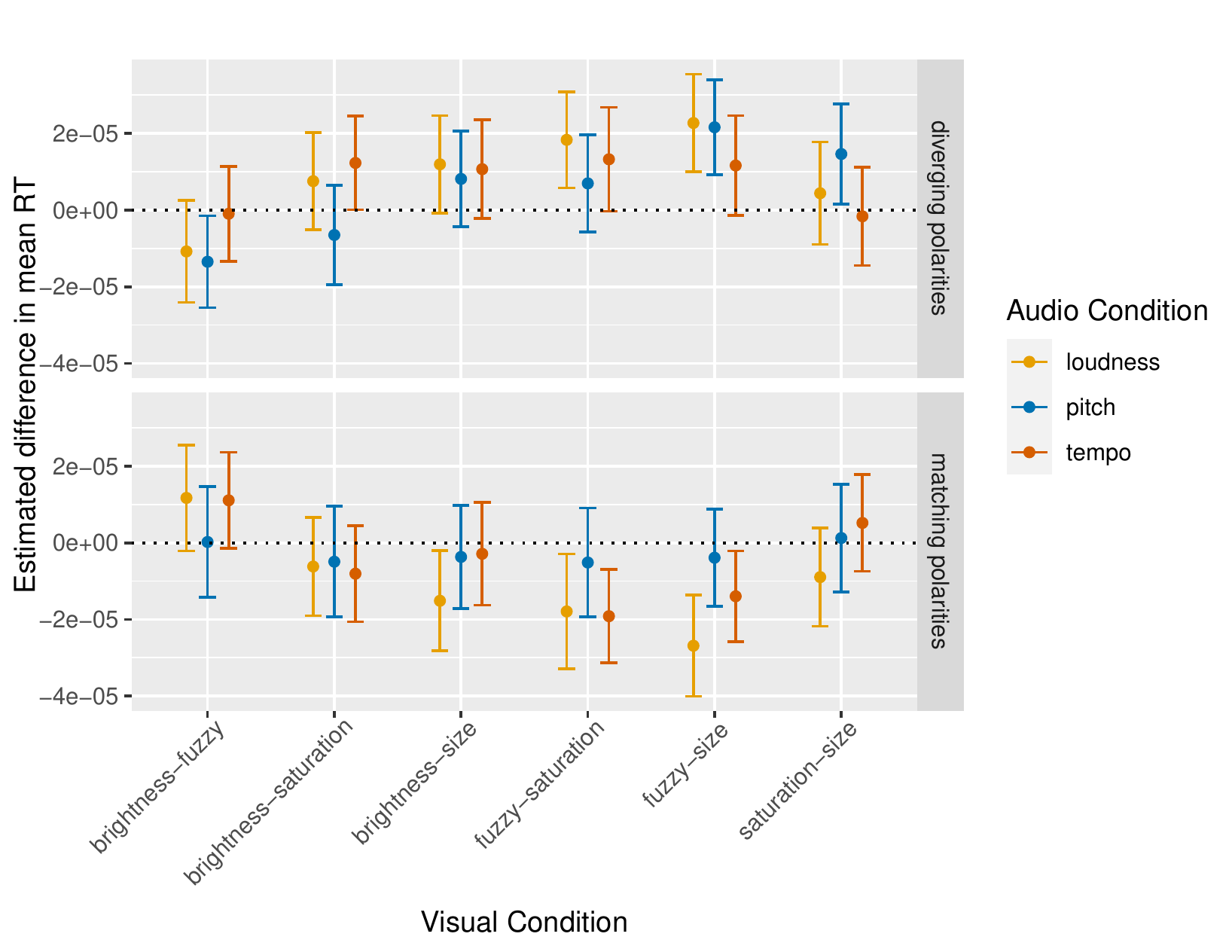}
\caption{The estimated difference of transformed mean RT of each AV pair per polarity mapping, tested between each visual condition, at the audio condition and polarity mapping level. The error bars show the 95\% CI. The further away from the reference line (0) the CI is, the bigger the difference in RT between the pairs.}
\Description{The image shows a series of 95\% CI plots of the pairwise analysis results. The x-axis represents the visual condition pairs being compared, such as brightness-fuzzy, brightness-saturation, and fuzzy-size. The y-axis represents the estimate. Each 95\% CI in the graph represents a different audio condition, which are pitch, loudness, and tempo. The graphs are organized into two rows, one for the matching polarities condition and one for diverging polarity condition. The overall trend of the graphs is that the estimate tends to be higher for matching polarities trials than for the diverging polarity condition. }
\label{fig:exp2_ci_pair}
\end{figure}

% prep the data
We analyzed RT according to the hypothesis: \textbf{There is an effect of audio channel, visual channel, and AV polarity mapping on reaction time (RT).} The Shapiro-Wilk test showed that the RT data significantly deviated from a normal distribution ($p < 0.001$). After transforming the RT data by dividing one over RT (1/RT), the data followed a normal distribution.

% analysis
We performed a three-way repeated measures ANOVA (audio channel $\times$ visual channel $\times$ polarity mapping). While we did not find enough evidence for the hypothesis ($F(6, 1104)=1.43, p=0.20$), we saw a significant effect of the interaction between visual channel $\times$ polarity mapping ($F(3, 552) = 13.24, p < 0.001$, $\eta_{p}^{2} = 0.067$), and of the audio condition ($F(2, 368) = 6.10, p = 0.032, \eta_{p}^{2} = 0.032$). 

%  visual channel X AV congruency at each SND
We investigated the interaction effect of visual channel $\times$ polarity mapping. A two-way ANOVA with Bonferroni correction showed a significant interaction on loudness ($F(3, 552) = 9.79, p < 0.001$, $\eta_{p}^{2} = 0.051$), as well as on tempo ($F(3, 552) = 5.33, p = 0.01, \eta_{p}^{2} = 0.028$).

% snd condition X  vis condition at each av congruency
Additionally, we carried out a two-way Bonferroni-corrected ANOVA to investigate the interaction effect between audio condition $\times$ visual condition at each polarity mapping. We saw a significance of AV pairs with matching polarities on the visual condition ($F(3, 552) = 5.75, p = 0.00, \eta_{p}^{2} = 0.030$), and of diverging AV pairs on the visual condition ($F(3, 552) = 8.55, p < 0.001, \eta_{p}^{2} = 0.044$).  

% pair wise & descriptives
Despite the lack of statistical significance, we carried out a pairwise analysis. 
\Cref{fig:exp2_ci_pair} shows no significant difference in mean RT between most AV stimuli since the 95\% CI includes 0. We highlight \Cref{fig:exp2_ci_raw}, which shows the mean RT and 95\% CI for each AV stimulus per polarity mapping. \embraced{Loudness\polarup \polarup size} shows the lowest RT. In general, the RT is low for matching polarities of size and saturation paired with any audio dimensions, and for diverging polarities of brightness and fuzziness. The 95\% CI are also shorter in these cases, indicating more consistent performance.

\subsubsection{Mapping Strategies}
% task difficulty
\begin{figure}[h!]
\centering
\includegraphics[width=0.8\columnwidth]{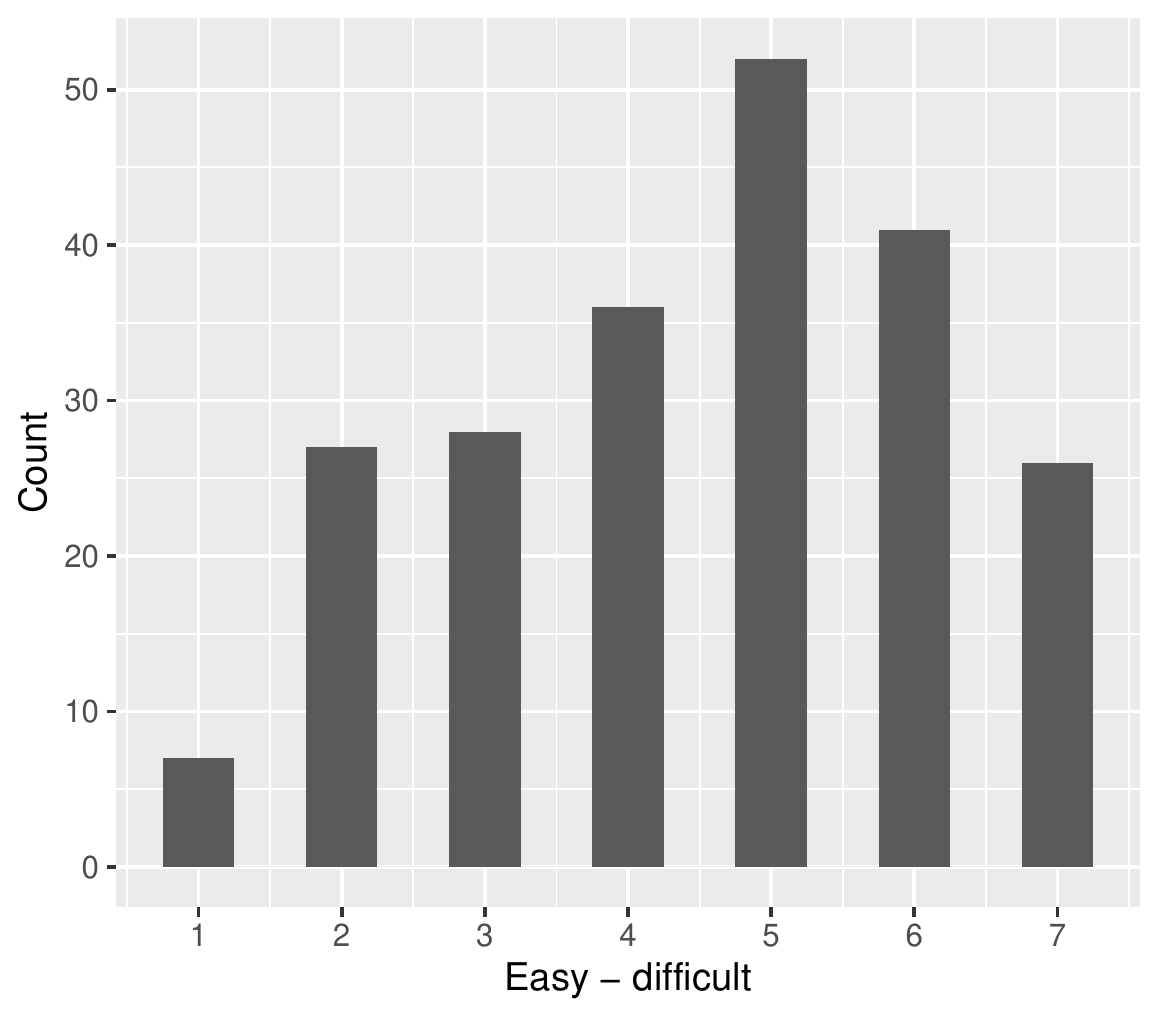}
\caption{A histogram of the difficulty of the task in Experiment~1, according to ratings by participants, on a 7-point Likert scale (from 1 = ``easy''  to 7 =  ``difficult''). The overall shape of the histogram is slightly skewed to the right, indicating participants thought the task was  mostly intermediate to difficult.
}
\Description{The figure is a histogram showing the distribution of responses on a scale from 1 to 7, representing levels of difficulty from easy to difficult. The x-axis represents the scale values from 1 to 7, while the y-axis represents the count of responses for each value. The overall shape of the histogram is skewed to the right, indicating that most responses are on the more difficult end of the scale. There is a clear peak at the value of 5, suggesting that this is the most common response.}
\label{fig:exp2_diff}
\end{figure}

% difficulty
\Cref{fig:exp2_diff} shows that participants found the task mostly intermediate to difficult. Participants found this experiment more difficult than Experiment~1, as indicated by comparing \Cref{fig:exp2_diff} to \Cref{fig:exp1_diff}.  

% themes
Participants mentioned similar strategies as they used in Experiment~1. We also discovered some strategies unique to the particular task from this experiment.  

% same as exp1
We found that 13 participants mentioned that they used a similar or the \textbf{same strategy} as they did in Experiment~1. For example, this participant wrote: \textit{``I applied the same logic as before [...]. I imagined one of those airport scanners in terms of sound. With visuals, I imagined a classic graph or heatmap.''}

% stim exposure
Another strategy was to \textbf{relate to previous stimuli exposure}. Essentially, 7 participants remembered the stimuli presented during a particular trial. Most participants remembered the stimuli from Experiment~1, like this participant: \textit{``I tried recalling experiences from the previous part of the experiment.''} One participant made the effort to memorize the stimuli: \textit{``[I] tried to memorize the sounds and~shapes.''}

% seperate
However, most participants \textbf{evaluated auditory and visual stimuli separately} (69). One participant wrote: \textit{``My strategy was to compare the sounds and images separately so I do not confuse the two or relate them to each other. Then I tried to connect low pitch with lower probability and high pitch with higher probability and try to make a pattern in my head about how it will go.''} Some participants mentioned that they mapped audio to probability and visual to probability, like this participant: \textit{``The bigger the shape, the higher the probability. The brighter the color, the higher the probability. The louder the sound, the higher the probability and vice versa.''}

% holistic
In contrast, 19 participants \textbf{evaluated auditory and visual stimuli as one}. Some wrote that they looked for common features in the audio and visual stimuli: \textit{``If [AV stimulus] B had sounds and images that were more alive I considered them higher probability.''} Another participant mentioned they focused on the AV stimuli first, before considering probability: \textit{``While hearing and seeing stimuli A I imagined what could happen next and if it was something similar which matches my idea I [chose] high probability.''}

% audio
We found 13 participants who \textbf{mainly focused on auditory stimulus attributes}. One of them mentioned: \textit{``Depending on the sound more [than on] the picture.''}

% visual
Fewer participants (8) \textbf{mainly focused on visual stimulus attributes.} For example: \textit{``I just based it on what I thought was accurate visually.''}

% audio-visual mapping
It seemed that 20 participants \textbf{mapped audio-visual first} before mapping to probability. One participant mentioned that they compared the stimulus to the anchor: \textit{``I tried to evaluate colours and sounds in connection with stimuli A and the same like in [...] task one. [The] highest sound made connection in my head with brighter colour or smaller circle.''} Some of these participants looked for common features between the auditory and visual stimuli first: \textit{``Well, I tried to find some common features in both stimulus options and evaluate them based on that.''}

% varying
One participant's \textbf{strategy changed throughout the experiment:} \textit{``Rereading the instructions multiple times but then it just clicked after few trials. I've used the previous strategy [from Experiment~1] but I did shift attention toward the beeping pitch a bit more.''} This was a strategy we also discovered in Experiment~1.

% intuition
We found 19 participants who mentioned that they relied on their \textbf{intuition}, a strategy we also noticed in Experiment~1. This participant explicitly mentioned they relied on intuition like in Experiment~1: \textit{``The same thing from [Experiment~1], just what feels right to me.''}

% unsure
We discovered that 11 participants were \textbf{unsure}. Some participants might have found the task difficult and could not focus on the AV stimulus-probability mapping: \textit{``This was truly difficult to make a pattern. I used the same mental judgment on matching sounds [as in Experiment~1] but I was not sure how to judge the image vs the sounds [...].''}

\subsection{Discussion}
% mapping pref
We saw strong agreements in Experiment~2. Particularly strong mappings included \embraced{loudness\polarup \polarup saturation}, \embraced{loudness\polarup \polarup size},  %\embraced{pitch\polarup\polarup saturation}, 
\inlinemaths{\langle}pitch\polarup \mbox{} \polarup  saturation\inlinemaths{\rangle},
\embraced{tempo\polarup\polarup saturation}, and \embraced{pitch\polarup \polarup size}. 
%
% Sorted entries from the table in the appendix, as basis of the above:
%loudness positive saturation positive 0.604
%loudness positive size positive 0.603
%frequency positive saturation positive 0.576 
%tempo positive saturation positive 0.568
%frequency positive size positive 0.551
%tempo positive size positive 0.542 
%loudness positive fuzzy negative 0.539 
%tempo positive fuzzy negative 0.515
%frequency positive fuzzy negative 0.514
%
An increase in the audio channel combined with an increase in the visual channel represents an increase in probability for these mappings. A decrease in probability is represented by a decrease in visual and audio dimensions. The mappings \embraced{loudness\polarup \polardown fuzziness}, 
\inlinemaths{\langle}tempo\polarup \mbox{} \polardown  fuzziness\inlinemaths{\rangle},
and \embraced{pitch\polarup \polardown fuzziness} were also preferred. For these mappings, an increase in the audio dimension and a decrease in the visual channel maps to an increasing probability. In contrast, for example, a decrease in probability is represented by a decrease in tempo and an increase in fuzziness.

When analyzing RT, we not only looked at the AV stimuli but also investigated the influence of the polarity mapping of the audio and visual dimensions of the stimuli. \Cref{fig:exp2_ci_raw} shows that when an AV stimulus with matching polarities leads to low RT, the AV stimulus with diverging polarities leads to a high RT and vice versa. Additionally, we noticed that the 95\% CI of AV pairs with low RT are much shorter, indicating more precise results.

% commonality
Most of the preferred AV mappings also resulted in the quickest responses, making these the best candidates to represent uncertainty. These mappings likely facilitated the mapping task, similar to the speeded classification paradigm \cite{marks2004cross}. These included \embraced{loudness\polarup \polarup size}, \embraced{loudness\polarup\polarup saturation}, \embraced{tempo\polarup \polarup saturation}, and \embraced{loudness\polarup \polardown fuzziness}. 

% discrepancy
However, some less preferred mappings resulted in low RT. \embraced{Loudness\polarup \polardown brightness} leads to one of the quickest RT while not being a highly preferred mapping. This is also true for \inlinemaths{\langle}tempo\polarup \mbox{} \polardown  brightness \inlinemaths{\rangle}. In contrast, not all of the most preferred AV-probability mappings resulted in low RT, such as \embraced{pitch\polarup \polarup saturation}, \inlinemaths{\langle}tempo\polarup \mbox{} \polarup size \inlinemaths{\rangle}, and \embraced{pitch\polarup \polarup size}.

We draw our conclusion from \Cref{fig:exp2_ci_raw} and \Cref{fig:exp2_ci_pair}, since we were unable to reject $H0$: There is an effect of audio channel, visual channel, and AV polarity mapping on reaction time (RT). Moreover, a majority of participants reported using a strategy to keep the audio and visual dimensions separate, and that the task was rather difficult.

%%%%%%%%%%%% Discussion and conclusion %%%%%%%%%%%%%%%

\section{General Discussion}

We first discuss the results of the experiments from an overall perspective. Later, we arrive at implications for future designs of AV representations of uncertainty.

\subsection{Aggregated Discussion of Experiment Results}
% strong preferences
We found AV pairs (Experiment~1) and AV-probability mappings (Experiment~2) with strong preferences. These preferences seem much stronger than data-to-audio mappings found in earlier sonification research \cite{walker2005mappings}. This could be due to the nature of the data mapped to the stimuli. However, while some earlier research into audio mappings for uncertainty found strong polarity preferences \cite{batterman2012auditory, batterman2013auditory}, other research had mixed results \cite{ballatore2019sonifying}. It may be the combination of senses that sends a stronger signal. Integrating sonification and visualization might represent uncertainty better than sonification or visualization alone.    

% discrepancy exp1 exp2 and 
However, not all preferred AV pairings are suitable for encoding uncertainty. While \embraced{pitch\polarup \polarup brightness} is a highly preferred pair, it is not a good probability mapping. In contrast, \embraced{tempo\polarup \polarup size} does not seem like a suitable AV pair but is a highly preferred mapping for probability. Experiment~2 is most relevant since it directly explores AV mappings of probability. Experiment~1 serves as a control to test whether AV pairs studied by cross-modal research are directly usable as AV-data mappings. Our results suggest that good AV pairs are not necessarily good probability mappings. Furthermore, it is unlikely that the results of Experiment~2 are due to ease of processing since the results of the experiments differ. Instead, users do not seem to rely solely on the AV pair when mapping to data concepts. This highlights that researchers should carefully consider AV-to-data mapping when designing AV analytics.

% loudness note
A better strategy when considering AV-to-data mapping is to pair audio and visual dimensions that are good mappings of the given data value in sonification and visualization research. Our results of AV mapping of uncertainty overlap with audio channels considered suitable for sonifying uncertainty and visual channels considered suitable for visualizing uncertainty. Like earlier research, loudness was especially suitable for AV semiotics of uncertainty~\cite{ballatore2019sonifying}. Likewise, fuzziness seems a suitable visual dimension in our research, similar to previous research~\cite{maceachren2012visual}. However, fuzziness does not dominate like loudness. Perhaps audio is an especially effective channel for conveying uncertainty. 

However, it is worth noting that the experimental setup of earlier research differs from ours. This is partially due to the nature of the stimuli: auditory stimuli can only be presented temporally, whereas visual stimuli can be presented spatially. It should also be noted that different studies used different incarnations of the broad concept of uncertainty. In our research, mappings were made to probability. In contrast, MacEachren et al.~\cite{maceachren2012visual} asked participants to map to uncertainty concepts such as accuracy and trustworthiness, and Ballatore et al.~\cite{ballatore2019sonifying} asked participants to map to data quality.

% guidelines
\subsection{Design Implications} \label{design}

\begin{figure}[b]
\centering
\includegraphics[width=\columnwidth]{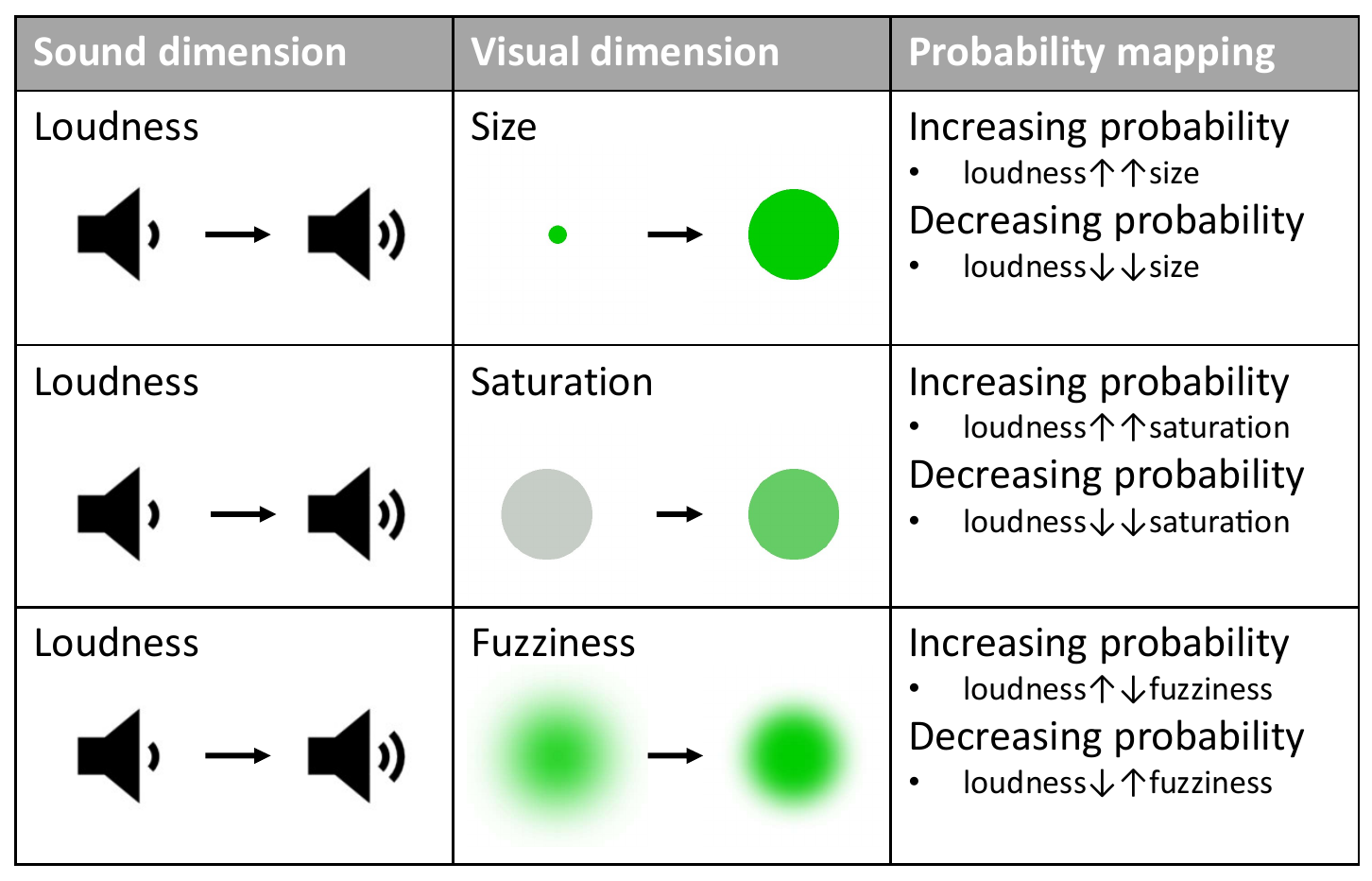}
\caption{The AV pairs we recommend for representing uncertainty based on the results from Experiment~1 and 2. All of these AV combinations represent an increase in probability.}
\Description{The figure shows a table with the recommended AV stimulus dimensions for representing uncertainty. The table consists of three columns: Sound dimension, Visual dimension, and Probability mapping. The first row under the column titles shows, from left to right: two icons that represent increasing loudness, two icons that represent increasing size, a text description signifying that the AV stimulus is matching polarities since they both increase. In the next row, two icons represent increasing loudness, two icons represent increasing saturation, and a text description signifying that the AV stimulus is matching polarities since they both increase. In the next row, two icons represent increasing loudness, two icons represent decreasing fuzziness, a text description signifying that the AV stimulus is diverging polarities since one dimension increases while the other decreases.}
\label{fig:res_fig}
\end{figure}

% design decisions
Researchers interested in AV semiotics of uncertainty should take into account the following design choices: \textbf{1) The audio and visual channel to pair.} It is best to choose an AV pair with a high preference according to our results, such as \embraced{loudness/size}, which ensures the mapping is considered intuitive to a majority of users. \textbf{2) The polarity matching of the AV pairs.} Our results show a significant interaction of polarity mapping, which suggests that a mapping can only be considered good if the polarity matching is considered good. For example,  
\embraced{loudness\polarup \polarup size} is well suited for increasing probability, whereas \embraced{loudness\polarup \polardown size} is not. \textbf{3) The mapping of AV dimension to data concept.} For example, while our findings show that \embraced{pitch\polarup \polarup brightness} is a preferred pair, it is not a good mapping for probability. 

% overlap
The results from both experiments should be carefully considered to make an informed decision. In some cases, we found only small differences in performance and preference. The difference between \embraced{pitch\polarup \polardown size} and \embraced{tempo\polarup \polardown size} is negligible when comparing the 95\% CI in \Cref{fig:exp2_ci_raw}. Both of these mappings could be considered equally good choices for representing probability.
However, there is a sizable difference between the mean RT of \embraced{loudness\polarup \polarup fuzziness} and \embraced{loudness\polarup \polarup size}. In this case, it is better to choose the mapping that leads to the higher preference: \embraced{loudness\polarup \polarup size}. In general, we recommend the mappings in \Cref{fig:res_fig} as they led to a high preference and low RT in both experiments.

% loudness and perceptual issues
While loudness is suitable according to our results, the dimension might not be the most robust. According to Hermann et al.~\cite{hermann2011sonification}, loudness is a poor choice for sonification since human memory of loudness is poor. Instead, pitch could be suitable since the human auditory system can detect very small changes in pitch \cite{neuhoff2019sonification}. Mappings such as \embraced{pitch\polarup \polarup saturation} are also found highly suitable.

\section{Limitations and Future Work}
% ecological validity
% log-rithmic
We did not account for the logarithmic nature of perception \cite{portugal2011weber} since we focused on polarity. However, participants were able to perceive differences between all anchors and stimuli with high accuracy during our visual and audio comparison checks (Appendix~\ref{clr_task} and Appendix~\ref{snd_task}), and participants were highly consistent in their polarity preferences (Appendix~\ref{exp1_consistency}). We believe that our results can be extended to communicate a range of probabilities, such as a probability density function (PDF). Readers interested in integrating AV semiotics into PDFs should take into account that users perceive changes in stimuli logarithmically and should investigate scaling. A logarithmic scaling approach, such as applying Weber's law \cite{portugal2011weber, goldstein2009encyclopedia, pringle1911Weber} would be a good start. 
We encourage future research to reuse the scripts we used to generate the AV stimuli to investigate the magnitude channels in applied settings with a multi-sensory data representation \cite{Vriend:2025:DarusSupplemental}.

Some users might want to compare data values. Error bars, for example, allow users to compare uncertainty by relating the length of the error bars. In such a case, users are mainly interested in understanding which plot presents the least (or most uncertainty). Our findings are well suited for such cases. For instance, we want to compare different models predicting the weather tomorrow. The plot's saturation and loudness map the model's probability to \embraced{loudness\polarup \polarup saturation}: the model with a more probable prediction is represented by a higher loudness and saturation compared to the less probable prediction.

% generalizability and semantics
We asked participants to map stimuli to high and low probability. However, it is impossible to control exactly how participants evaluated the stimuli, especially in a crowd-sourced study. Walker~\cite{walker2007consistency} points out that the semantics of data variables can affect polarity mapping in sonification. This means that our explanation of probability and of ``high probability'' and ``low probability'' could have affected mapping choices. Future research should investigate to what extent our results are generalizable to other uncertainty concepts. Also, it is possible that participants did not map the stimuli to high or low probability but to something more general, regardless of data concept, for example ``high'' and ``low.'' In visualization, guidelines exist that apply across data concepts. For example, the ``dark-is-more'' principle maps increased darkness to higher data values, regardless of color or measure \cite{Schloss2019colormap, Schoenlein2023relational}. This could extend to the AV dimension too, making the mappings generalizable. Future research should examine how semantics affects AV-data mappings. 

% individual differences 
None of the AV stimuli we tested resulted in a unanimous preference. Additionally, we discovered a wide range of mapping strategies, which could be due to personal differences. While some sonification research on differences in musicality found mixed results \cite{lacherez2007overlapping, sandor2003sonification}, not much research has been done on other factors such as cultural, perceptual, or cognitive differences \cite{Walker2023promise, nees2007listener}. Possible avenues for future research could explore how mapping strategies and performance are affected by, for example, perceptual differences. Results from such studies could help predict the kind of mappings users prefer based on personal differences. We should point out, however, that participants reported their mapping strategy for all trials per experiment. Participants likely developed different mental models per pair or mapping. Preferred pairs and mappings might be processed holistically, while least preferred pairs and mappings might be processed separately. Future research should investigate these mental models of AV semiotics further. 

% universality
Finally, it is impossible to examine the complete AV design space. Hence, our semiotics of uncertainty is not all-encompassing. We chose a subset of dimensions from previous work \cite{ballatore2019sonifying, maceachren2012visual}. Other audio and visual channels from these studies that do not represent uncertainty well on their own might work well when combined. Additionally, channels such as spatial sound could be explored.

\section{Conclusion}
% summary
This research contributes to the limited research in AV data representation \cite{enge2023towards, enge2024open} by exploring AV semiotics of uncertainty. Experiment~1 explored AV pairs and showed high preferences and low RT for 
\embraced{loudness\polarup \polarup size}, \embraced{loudness\polarup \polarup saturation},  \embraced{pitch\polarup \polarup brightness}, 
and \embraced{loudness\polarup \polardown fuzziness}.
% Old version / partially wrong, and often with wrong order:
%\embraced{brightness\polarup\polarup pitch}, %\embraced{loudness\polarup\polarup saturation}, \\
%\embraced{loudness\polarup\polarup saturation}, and 
%\embraced{fuzziness\polarup\polardown loudness}. 
Experiment~2 explored AV mappings of probability. We found low RT and high preference for %\embraced{loudness\polarup \polarup saturation}, 
\inlinemaths{\langle}loudness\polarup \mbox{} \polarup  saturation\inlinemaths{\rangle},
\embraced{loudness\polarup \polarup size}, 
\embraced{pitch\polarup \polarup saturation}, \embraced{tempo\polarup \polarup sat\-uration}, 
\embraced{tempo\polarup \polarup size}, and 
\embraced{loudness\polarup \polardown fuzziness}.

% Preference table:
%loudness positive saturation positive 0.604
%loudness positive size positive 0.603
%frequency positive saturation positive 0.576 
%tempo positive saturation positive 0.568
%frequency positive size positive 0.551
%tempo positive size positive 0.542 
%loudness positive fuzzy negative 0.539 
%tempo positive fuzzy negative 0.515
%frequency positive fuzzy negative 0.514

Additionally, we explored the mapping strategies participants used in both experiments to contribute to the limited research on how users translate between stimulus channel and data concept~\cite{walker2007consistency}. We found a wide variety of strategies, some overlapping between the two experiments.

We showed that preferred AV pairs do not necessarily make suitable mappings. Future research integrating visualization and sonification should not only consider the AV channels but also whether the AV channels map to their intended data concept. For uncertainty, we recommend AV channels that are both good AV pairs and good mappings.

 \begin{acks}
      The first author would like to thank Marc O. Ernst and Priscilla Balestrucci for their discussions and insights at the beginning of this research. The authors thank the participants for their time and effort. This work was funded by the 
\grantsponsor{DFG}{Deutsche Forschungsgemeinschaft (DFG, German Research Foundation)}{https://www.dfg.de}---Project-ID \grantnum{DFG}{251654672}---TRR~161.
\end{acks}

\bibliographystyle{ACM-Reference-Format}
\bibliography{AM25/revised_ref}


\begin{thebibliography}{76}

%%% the Web address.

\ifx \showCODEN    \undefined \def \showCODEN     #1{\unskip}     \fi
\ifx \showDOI      \undefined \def \showDOI       #1{#1}\fi
\ifx \showISBNx    \undefined \def \showISBNx     #1{\unskip}     \fi
\ifx \showISBNxiii \undefined \def \showISBNxiii  #1{\unskip}     \fi
\ifx \showISSN     \undefined \def \showISSN      #1{\unskip}     \fi
\ifx \showLCCN     \undefined \def \showLCCN      #1{\unskip}     \fi
\ifx \shownote     \undefined \def \shownote      #1{#1}          \fi
\ifx \showarticletitle \undefined \def \showarticletitle #1{#1}   \fi
\ifx \showURL      \undefined \def \showURL       {\relax}        \fi
% invisible to TeX
\providecommand\bibfield[2]{#2}
\providecommand\bibinfo[2]{#2}
\providecommand\natexlab[1]{#1}
\providecommand\showeprint[2][]{arXiv:#2}

\bibitem[Aigner et~al\mbox{.}(2022)]%
        {aigner2022workshop}
\bibfield{author}{\bibinfo{person}{Wolfgang Aigner}, \bibinfo{person}{Kajetan
  Enge}, \bibinfo{person}{Michael Iber}, \bibinfo{person}{Alexander Rind},
  \bibinfo{person}{Niklas Elmqvist}, \bibinfo{person}{Robert H\"{o}ldrich},
  \bibinfo{person}{Niklas R\"{o}nnberg}, {and} \bibinfo{person}{Bruce~N.
  Walker}.} \bibinfo{year}{2022}\natexlab{}.
\newblock \showarticletitle{Workshop on audio-visual analytics}. In
  \bibinfo{booktitle}{\emph{Proceedings of the 2022 International Conference on
  Advanced Visual Interfaces}} (Frascati, Rome, Italy)
  \emph{(\bibinfo{series}{AVI 2022})}. \bibinfo{publisher}{Association for
  Computing Machinery}, \bibinfo{address}{New York, NY, USA}, Article
  \bibinfo{articleno}{92}, \bibinfo{numpages}{4}~pages.
\newblock
\showISBNx{9781450397193}
\urldef\tempurl%
\url{https://doi.org/10.1145/3531073.3535252}
\showDOI{\tempurl}


\bibitem[Anwar et~al\mbox{.}(2023)]%
        {Anwar2023haptic}
\bibfield{author}{\bibinfo{person}{Ahmed Anwar}, \bibinfo{person}{Tianzheng
  Shi}, {and} \bibinfo{person}{Oliver Schneider}.}
  \bibinfo{year}{2023}\natexlab{}.
\newblock \showarticletitle{Factors of haptic experience across multiple haptic
  modalities}. In \bibinfo{booktitle}{\emph{Proceedings of the 2023 CHI
  Conference on Human Factors in Computing Systems}} (Hamburg, Germany)
  \emph{(\bibinfo{series}{CHI})}. \bibinfo{publisher}{Association for Computing
  Machinery}, \bibinfo{address}{New York, NY, USA}, Article
  \bibinfo{articleno}{260}, \bibinfo{numpages}{12}~pages.
\newblock
\showISBNx{9781450394215}
\urldef\tempurl%
\url{https://doi.org/10.1145/3544548.3581514}
\showDOI{\tempurl}


\bibitem[Ballatore et~al\mbox{.}(2018)]%
        {ballatore2019sonifying}
\bibfield{author}{\bibinfo{person}{Andrea Ballatore}, \bibinfo{person}{David
  Gordon}, {and} \bibinfo{person}{Alexander~P. Boone}.}
  \bibinfo{year}{2018}\natexlab{}.
\newblock \showarticletitle{Sonifying data uncertainty with sound dimensions}.
\newblock \bibinfo{journal}{\emph{Cartography and Geographic Information
  Science}} \bibinfo{volume}{46}, \bibinfo{number}{5} (\bibinfo{year}{2018}),
  \bibinfo{pages}{385--400}.
\newblock
\urldef\tempurl%
\url{https://doi.org/10.1080/15230406.2018.1495103}
\showDOI{\tempurl}


\bibitem[Batterman and Walker(2012)]%
        {batterman2012auditory}
\bibfield{author}{\bibinfo{person}{Jared~M. Batterman} {and}
  \bibinfo{person}{Bruce~N. Walker}.} \bibinfo{year}{2012}\natexlab{}.
\newblock \showarticletitle{Displaying error \& uncertainty in auditory
  graphs}. In \bibinfo{booktitle}{\emph{Proceedings of the 14th International
  ACM SIGACCESS Conference on Computers and Accessibility}} (Boulder, Colorado,
  USA) \emph{(\bibinfo{series}{ASSETS})}. \bibinfo{publisher}{Association for
  Computing Machinery}, \bibinfo{address}{New York, NY, USA},
  \bibinfo{pages}{285–286}.
\newblock
\showISBNx{9781450313216}
\urldef\tempurl%
\url{https://doi.org/10.1145/2384916.2384995}
\showDOI{\tempurl}


\bibitem[Batterman and Walker(2013)]%
        {batterman2013auditory}
\bibfield{author}{\bibinfo{person}{Jarred~M. Batterman} {and}
  \bibinfo{person}{Bruce~N. Walker}.} \bibinfo{year}{2013}\natexlab{}.
\newblock \showarticletitle{Auditory graphs need error bars: Validating
  error-to-sound mappings and scalings}. In
  \bibinfo{booktitle}{\emph{Proceedings of the International Conference on
  Auditory Display}} (\L{}\'{o}d\'{z}, Poland) \emph{(\bibinfo{series}{ICAD})}.
  \bibinfo{pages}{315--318}.
\newblock


\bibitem[Bearman(2011)]%
        {bearman2011using}
\bibfield{author}{\bibinfo{person}{Nick Bearman}.}
  \bibinfo{year}{2011}\natexlab{}.
\newblock \showarticletitle{Using sound to represent uncertainty in future
  climate predictions for the {United Kingdom}}. In
  \bibinfo{booktitle}{\emph{Proceedings of the International Conference on
  Auditory Display}} (Budapest, Hungary) \emph{(\bibinfo{series}{ICAD})}.
  \bibinfo{publisher}{Georgia Institute of Technology}.
\newblock


\bibitem[Belia et~al\mbox{.}(2005)]%
        {belia2005researchers}
\bibfield{author}{\bibinfo{person}{Sarah Belia}, \bibinfo{person}{Fiona
  Fidler}, \bibinfo{person}{Jennifer Williams}, {and} \bibinfo{person}{Geoff
  Cumming}.} \bibinfo{year}{2005}\natexlab{}.
\newblock \showarticletitle{Researchers misunderstand confidence intervals and
  standard error bars}.
\newblock \bibinfo{journal}{\emph{Psychological Methods}} \bibinfo{volume}{10},
  \bibinfo{number}{4} (\bibinfo{year}{2005}), \bibinfo{pages}{389--396}.
\newblock
\urldef\tempurl%
\url{https://doi.org/10.1037/1082-989X.10.4.389}
\showDOI{\tempurl}


\bibitem[Bertin(1983)]%
        {bertin1983semiology}
\bibfield{author}{\bibinfo{person}{J. Bertin}.}
  \bibinfo{year}{1983}\natexlab{}.
\newblock \bibinfo{booktitle}{\emph{Semiology of Graphics: Diagrams, Networks,
  Maps}}.
\newblock \bibinfo{publisher}{University of Wisconsin Press},
  \bibinfo{address}{Madison, WI, USA}.
\newblock
\showISBNx{0299090604}


\bibitem[Bonneau et~al\mbox{.}(2014)]%
        {Bonneau:2014:OSU}
\bibfield{author}{\bibinfo{person}{Georges-Pierre Bonneau},
  \bibinfo{person}{Hans-Christian Hege}, \bibinfo{person}{Chris~R. Johnson},
  \bibinfo{person}{Manuel~M. Oliveira}, \bibinfo{person}{Kristin Potter},
  \bibinfo{person}{Penny Rheingans}, {and} \bibinfo{person}{Thomas Schultz}.}
  \bibinfo{year}{2014}\natexlab{}.
\newblock \showarticletitle{Overview and state-of-the-art of uncertainty
  visualization}.
\newblock In \bibinfo{booktitle}{\emph{Scientific Visualization: Uncertainty,
  Multifield, Biomedical, and Scalable Visualization}},
  \bibfield{editor}{\bibinfo{person}{Charles~D. Hansen}, \bibinfo{person}{Min
  Chen}, \bibinfo{person}{Christopher~R. Johnson}, \bibinfo{person}{Arie~E.
  Kaufman}, {and} \bibinfo{person}{Hans Hagen}} (Eds.).
  \bibinfo{publisher}{Springer}, \bibinfo{address}{London, UK},
  \bibinfo{pages}{3--27}.
\newblock
\showISBNx{978-1-4471-6497-5}
\urldef\tempurl%
\url{https://doi.org/10.1007/978-1-4471-6497-5_1}
\showDOI{\tempurl}


\bibitem[Boukhelifa et~al\mbox{.}(2012)]%
        {boukhelifa2012sketchiness}
\bibfield{author}{\bibinfo{person}{Nadia Boukhelifa},
  \bibinfo{person}{Anastasia Bezerianos}, \bibinfo{person}{Tobias Isenberg},
  {and} \bibinfo{person}{Jean-Daniel Fekete}.} \bibinfo{year}{2012}\natexlab{}.
\newblock \showarticletitle{Evaluating sketchiness as a visual variable for the
  depiction of qualitative uncertainty}.
\newblock \bibinfo{journal}{\emph{IEEE Transactions on Visualization and
  Computer Graphics}} \bibinfo{volume}{18}, \bibinfo{number}{12}
  (\bibinfo{year}{2012}), \bibinfo{pages}{2769--2778}.
\newblock
\urldef\tempurl%
\url{https://doi.org/10.1109/TVCG.2012.220}
\showDOI{\tempurl}


\bibitem[Brodlie et~al\mbox{.}(2012)]%
        {Brodlie:2012:RUD}
\bibfield{author}{\bibinfo{person}{Ken Brodlie},
  \bibinfo{person}{Rodolfo~Allendes Osorio}, {and} \bibinfo{person}{Adriano
  Lopes}.} \bibinfo{year}{2012}\natexlab{}.
\newblock \showarticletitle{A review of uncertainty in data visualization}.
\newblock In \bibinfo{booktitle}{\emph{Expanding the Frontiers of Visual
  Analytics and Visualization}}, \bibfield{editor}{\bibinfo{person}{John Dill},
  \bibinfo{person}{Rae Earnshaw}, \bibinfo{person}{David Kasik},
  \bibinfo{person}{John Vince}, {and} \bibinfo{person}{Pak~Chung Wong}} (Eds.).
  \bibinfo{publisher}{Springer}, \bibinfo{address}{London, UK},
  \bibinfo{pages}{81--109}.
\newblock
\urldef\tempurl%
\url{https://doi.org/10.1007/978-1-4471-2804-5_6}
\showDOI{\tempurl}


\bibitem[Caiola et~al\mbox{.}(2022)]%
        {caiola2022audiovisual}
\bibfield{author}{\bibinfo{person}{Valentina Caiola}, \bibinfo{person}{Sara
  Lenzi}, {and} \bibinfo{person}{Dina Ricco}.} \bibinfo{year}{2022}\natexlab{}.
\newblock \showarticletitle{Audiovisual sonifications. A design map for
  multisensory integration in data representation}. In
  \bibinfo{booktitle}{\emph{DRS2022}} (Bilbao, Spain). \bibinfo{publisher}{DRS
  Digital Library}, \bibinfo{pages}{1--19}.
\newblock
\showISBNx{978-1-91229-457-2}
\urldef\tempurl%
\url{https://doi.org/10.21606/drs.2022.380}
\showDOI{\tempurl}


\bibitem[Chundury et~al\mbox{.}(2024)]%
        {niklas2024tvcg}
\bibfield{author}{\bibinfo{person}{Pramod Chundury}, \bibinfo{person}{Yasmin
  Reyazuddin}, \bibinfo{person}{J.~Bern Jordan}, \bibinfo{person}{Jonathan
  Lazar}, {and} \bibinfo{person}{Niklas Elmqvist}.}
  \bibinfo{year}{2024}\natexlab{}.
\newblock \showarticletitle{TactualPlot: Spatializing data as sound using
  sensory substitution for touchscreen accessibility}.
\newblock \bibinfo{journal}{\emph{IEEE Transactions on Visualization and
  Computer Graphics}} \bibinfo{volume}{30}, \bibinfo{number}{1}
  (\bibinfo{year}{2024}), \bibinfo{pages}{836--846}.
\newblock
\urldef\tempurl%
\url{https://doi.org/10.1109/TVCG.2023.3326937}
\showDOI{\tempurl}


\bibitem[Cooke et~al\mbox{.}(2017)]%
        {cooke2017exploring}
\bibfield{author}{\bibinfo{person}{Jeff Cooke}, \bibinfo{person}{W.
  D{\'\i}az-Merced}, \bibinfo{person}{Garry Foran}, \bibinfo{person}{Jeffrey
  Hannam}, {and} \bibinfo{person}{B. Garcia}.} \bibinfo{year}{2017}\natexlab{}.
\newblock \showarticletitle{Exploring data sonification to enable, enhance, and
  accelerate the analysis of big, noisy, and multi-dimensional data: workshop
  9}.
\newblock \bibinfo{journal}{\emph{Proceedings of the International Astronomical
  Union}} \bibinfo{volume}{14}, \bibinfo{number}{S339} (\bibinfo{year}{2017}),
  \bibinfo{pages}{251--256}.
\newblock
\urldef\tempurl%
\url{https://doi.org/10.1017/S1743921318002703}
\showDOI{\tempurl}


\bibitem[Correll et~al\mbox{.}(2018)]%
        {correll2018value}
\bibfield{author}{\bibinfo{person}{Michael Correll}, \bibinfo{person}{Dominik
  Moritz}, {and} \bibinfo{person}{Jeffrey Heer}.}
  \bibinfo{year}{2018}\natexlab{}.
\newblock \showarticletitle{Value-suppressing uncertainty palettes}. In
  \bibinfo{booktitle}{\emph{Proceedings of the 2018 CHI Conference on Human
  Factors in Computing Systems}} (Montreal QC, Canada)
  \emph{(\bibinfo{series}{CHI})}. \bibinfo{publisher}{Association for Computing
  Machinery}, \bibinfo{address}{New York, NY, USA}, Article
  \bibinfo{articleno}{642}, \bibinfo{numpages}{11}~pages.
\newblock
\urldef\tempurl%
\url{https://doi.org/10.1145/3173574.3174216}
\showDOI{\tempurl}


\bibitem[Dubus and Bresin(2013)]%
        {dubus2013systematic}
\bibfield{author}{\bibinfo{person}{Ga{\"e}l Dubus} {and}
  \bibinfo{person}{Roberto Bresin}.} \bibinfo{year}{2013}\natexlab{}.
\newblock \showarticletitle{A systematic review of mapping strategies for the
  sonification of physical quantities}.
\newblock \bibinfo{journal}{\emph{PLoS ONE}} \bibinfo{volume}{8},
  \bibinfo{number}{12} (\bibinfo{year}{2013}), \bibinfo{pages}{e82491}.
\newblock
\urldef\tempurl%
\url{https://doi.org/10.1371/journal.pone.0082491}
\showDOI{\tempurl}


\bibitem[Enge et~al\mbox{.}(2024)]%
        {enge2024open}
\bibfield{author}{\bibinfo{person}{K. Enge}, \bibinfo{person}{E. Elmquist},
  \bibinfo{person}{V. Caiola}, \bibinfo{person}{N. Rönnberg},
  \bibinfo{person}{A. Rind}, \bibinfo{person}{M. Iber}, \bibinfo{person}{S.
  Lenzi}, \bibinfo{person}{F. Lan}, \bibinfo{person}{R. Höldrich}, {and}
  \bibinfo{person}{W. Aigner}.} \bibinfo{year}{2024}\natexlab{}.
\newblock \showarticletitle{Open your ears and take a look: A state-of-the-art
  report on the integration of sonification and visualization}.
\newblock \bibinfo{journal}{\emph{Computer Graphics Forum}}
  \bibinfo{volume}{43}, \bibinfo{number}{3} (\bibinfo{year}{2024}),
  \bibinfo{pages}{e15114}.
\newblock
\urldef\tempurl%
\url{https://doi.org/10.1111/cgf.15114}
\showDOI{\tempurl}


\bibitem[Enge et~al\mbox{.}(2023)]%
        {enge2023towards}
\bibfield{author}{\bibinfo{person}{Kajetan Enge}, \bibinfo{person}{Alexander
  Rind}, \bibinfo{person}{Michael Iber}, \bibinfo{person}{Robert H\"{o}ldrich},
  {and} \bibinfo{person}{Wolfgang Aigner}.} \bibinfo{year}{2023}\natexlab{}.
\newblock \showarticletitle{Towards a unified terminology for sonification and
  visualization}.
\newblock \bibinfo{journal}{\emph{Personal Ubiquitous Computing}}
  \bibinfo{volume}{27}, \bibinfo{number}{5} (\bibinfo{year}{2023}),
  \bibinfo{pages}{1949–1963}.
\newblock
\showISSN{1617-4909}
\urldef\tempurl%
\url{https://doi.org/10.1007/s00779-023-01720-5}
\showDOI{\tempurl}


\bibitem[Evans and Treisman(2010)]%
        {Evans2010crossmodal}
\bibfield{author}{\bibinfo{person}{Karla~K. Evans} {and} \bibinfo{person}{Anne
  Treisman}.} \bibinfo{year}{2010}\natexlab{}.
\newblock \showarticletitle{Natural cross-modal mappings between visual and
  auditory features}.
\newblock \bibinfo{journal}{\emph{Journal of Vision}} \bibinfo{volume}{10},
  \bibinfo{number}{1} (\bibinfo{year}{2010}), \bibinfo{pages}{6, 1--12}.
\newblock
\showISSN{1534-7362}
\urldef\tempurl%
\url{https://doi.org/10.1167/10.1.6}
\showDOI{\tempurl}


\bibitem[Fagerlin et~al\mbox{.}(2005)]%
        {fagerlin2005reducing}
\bibfield{author}{\bibinfo{person}{Angela Fagerlin}, \bibinfo{person}{Catharine
  Wang}, {and} \bibinfo{person}{Peter Ubel}.} \bibinfo{year}{2005}\natexlab{}.
\newblock \showarticletitle{Reducing the influence of anecdotal reasoning on
  people's health care decisions: Is a picture worth a thousand statistics?}
\newblock \bibinfo{journal}{\emph{Medical Decision Making}}
  \bibinfo{volume}{25}, \bibinfo{number}{4} (\bibinfo{year}{2005}),
  \bibinfo{pages}{398--405}.
\newblock
\urldef\tempurl%
\url{https://doi.org/10.1177/0272989X05278931}
\showDOI{\tempurl}


\bibitem[Ferguson and Brewster(2017)]%
        {ferguson2017evaluation}
\bibfield{author}{\bibinfo{person}{Jamie Ferguson} {and}
  \bibinfo{person}{Stephen~A. Brewster}.} \bibinfo{year}{2017}\natexlab{}.
\newblock \showarticletitle{Evaluation of psychoacoustic sound parameters for
  sonification}. In \bibinfo{booktitle}{\emph{Proceedings of the 19th ACM
  International Conference on Multimodal Interaction}} (Glasgow, UK)
  \emph{(\bibinfo{series}{ICMI})}. \bibinfo{publisher}{Association for
  Computing Machinery}, \bibinfo{address}{New York, NY, USA},
  \bibinfo{pages}{120–127}.
\newblock
\showISBNx{9781450355438}
\urldef\tempurl%
\url{https://doi.org/10.1145/3136755.3136783}
\showDOI{\tempurl}


\bibitem[Ferguson and Brewster(2018)]%
        {Ferguson2018congruence}
\bibfield{author}{\bibinfo{person}{Jamie Ferguson} {and}
  \bibinfo{person}{Stephen~A. Brewster}.} \bibinfo{year}{2018}\natexlab{}.
\newblock \showarticletitle{Investigating perceptual congruence between data
  and display dimensions in sonification}. In
  \bibinfo{booktitle}{\emph{Proceedings of the 2018 CHI Conference on Human
  Factors in Computing Systems}} (Montreal, QC, Canada)
  \emph{(\bibinfo{series}{CHI})}. \bibinfo{publisher}{Association for Computing
  Machinery}, \bibinfo{address}{New York, NY, USA}, Article
  \bibinfo{articleno}{611}, \bibinfo{numpages}{9}~pages.
\newblock
\showISBNx{9781450356206}
\urldef\tempurl%
\url{https://doi.org/10.1145/3173574.3174185}
\showDOI{\tempurl}


\bibitem[Freeman(2020)]%
        {FREEMAN202066}
\bibfield{author}{\bibinfo{person}{Elliot~D. Freeman}.}
  \bibinfo{year}{2020}\natexlab{}.
\newblock \showarticletitle{Hearing what you see: Distinct excitatory and
  disinhibitory mechanisms contribute to visually-evoked auditory sensations}.
\newblock \bibinfo{journal}{\emph{Cortex}}  \bibinfo{volume}{131}
  (\bibinfo{year}{2020}), \bibinfo{pages}{66--78}.
\newblock
\showISSN{0010-9452}
\urldef\tempurl%
\url{https://doi.org/10.1016/j.cortex.2020.06.014}
\showDOI{\tempurl}


\bibitem[Galesic et~al\mbox{.}(2009)]%
        {galesic2009using}
\bibfield{author}{\bibinfo{person}{Mirta Galesic}, \bibinfo{person}{Rocio
  Garcia-Retamero}, {and} \bibinfo{person}{Gerd Gigerenzer}.}
  \bibinfo{year}{2009}\natexlab{}.
\newblock \showarticletitle{Using icon arrays to communicate medical risks:
  overcoming low numeracy.}
\newblock \bibinfo{journal}{\emph{Health Psychology}} \bibinfo{volume}{28},
  \bibinfo{number}{2} (\bibinfo{year}{2009}), \bibinfo{pages}{210--216}.
\newblock
\urldef\tempurl%
\url{https://doi.org/10.1037/a0014474}
\showDOI{\tempurl}


\bibitem[Gallace and Spence(2006)]%
        {gallace2006multisensory}
\bibfield{author}{\bibinfo{person}{Alberto Gallace} {and}
  \bibinfo{person}{Charles Spence}.} \bibinfo{year}{2006}\natexlab{}.
\newblock \showarticletitle{Multisensory synesthetic interactions in the
  speeded classification of visual size}.
\newblock \bibinfo{journal}{\emph{Perception \& Psychophysics}}
  \bibinfo{volume}{68}, \bibinfo{number}{7} (\bibinfo{year}{2006}),
  \bibinfo{pages}{1191--1203}.
\newblock
\urldef\tempurl%
\url{https://doi.org/10.3758/BF03193720}
\showDOI{\tempurl}


\bibitem[Goldstein(2010)]%
        {goldstein2009encyclopedia}
\bibfield{author}{\bibinfo{person}{E.~Bruce Goldstein}.}
  \bibinfo{year}{2010}\natexlab{}.
\newblock \bibinfo{booktitle}{\emph{Encyclopedia of Perception}}.
\newblock \bibinfo{publisher}{SAGE Publications, Inc.},
  \bibinfo{address}{Thousand Oaks, CA, USA}.
\newblock
\showISBNx{978-1-4129-4081-8}


\bibitem[Greenwald et~al\mbox{.}(1998)]%
        {greenwald1998measuring}
\bibfield{author}{\bibinfo{person}{Anthony~G. Greenwald},
  \bibinfo{person}{Debbie~E. McGhee}, {and} \bibinfo{person}{Jordan L.~K.
  Schwartz}.} \bibinfo{year}{1998}\natexlab{}.
\newblock \showarticletitle{Measuring individual differences in implicit
  cognition: The implicit association test}.
\newblock \bibinfo{journal}{\emph{Journal of Personality and Social
  Psychology}} \bibinfo{volume}{74}, \bibinfo{number}{6}
  (\bibinfo{year}{1998}), \bibinfo{pages}{1464--1480}.
\newblock
\urldef\tempurl%
\url{https://doi.org/10.1037/0022-3514.74.6.1464}
\showDOI{\tempurl}


\bibitem[Gschwandtnei et~al\mbox{.}(2016)]%
        {gschwandtnei2016temporal}
\bibfield{author}{\bibinfo{person}{Theresia Gschwandtnei},
  \bibinfo{person}{Markus B\"ogl}, \bibinfo{person}{Paolo Federico}, {and}
  \bibinfo{person}{Silvia Miksch}.} \bibinfo{year}{2016}\natexlab{}.
\newblock \showarticletitle{Visual encodings of temporal uncertainty: A
  comparative user study}.
\newblock \bibinfo{journal}{\emph{IEEE Transactions on Visualization and
  Computer Graphics}} \bibinfo{volume}{22}, \bibinfo{number}{1}
  (\bibinfo{year}{2016}), \bibinfo{pages}{539--548}.
\newblock
\urldef\tempurl%
\url{https://doi.org/10.1109/TVCG.2015.2467752}
\showDOI{\tempurl}


\bibitem[H\"agele et~al\mbox{.}(2022)]%
        {haegeleIt2022}
\bibfield{author}{\bibinfo{person}{David H\"agele}, \bibinfo{person}{Christoph
  Schulz}, \bibinfo{person}{Cedric Beschle}, \bibinfo{person}{Hannah Booth},
  \bibinfo{person}{Miriam Butt}, \bibinfo{person}{Andrea Barth},
  \bibinfo{person}{Oliver Deussen}, {and} \bibinfo{person}{Daniel Weiskopf}.}
  \bibinfo{year}{2022}\natexlab{}.
\newblock \showarticletitle{Uncertainty visualization: Fundamentals and recent
  developments}.
\newblock \bibinfo{journal}{\emph{it - Information Technology}}
  \bibinfo{volume}{64}, \bibinfo{number}{4--5} (\bibinfo{year}{2022}),
  \bibinfo{pages}{121–132}.
\newblock
\urldef\tempurl%
\url{https://doi.org/10.1515/itit-2022-0033}
\showDOI{\tempurl}


\bibitem[Han(2017)]%
        {Yoon2017multimodal}
\bibfield{author}{\bibinfo{person}{Yoon~Chung Han}.}
  \bibinfo{year}{2017}\natexlab{}.
\newblock \showarticletitle{California drought impact v2: Data visualization
  and sonification using advanced multimodal interaction}. In
  \bibinfo{booktitle}{\emph{Proceedings of the 2017 CHI Conference Extended
  Abstracts on Human Factors in Computing Systems}} (Denver, Colorado, USA)
  \emph{(\bibinfo{series}{CHI})}. \bibinfo{publisher}{Association for Computing
  Machinery}, \bibinfo{address}{New York, NY, USA},
  \bibinfo{pages}{1372–1377}.
\newblock
\showISBNx{9781450346566}
\urldef\tempurl%
\url{https://doi.org/10.1145/3027063.3052542}
\showDOI{\tempurl}


\bibitem[Hermann et~al\mbox{.}(2011)]%
        {hermann2011sonification}
\bibfield{author}{\bibinfo{person}{Thomas Hermann}, \bibinfo{person}{Andy
  Hunt}, {and} \bibinfo{person}{John Neuhoff}.}
  \bibinfo{year}{2011}\natexlab{}.
\newblock \bibinfo{booktitle}{\emph{The Sonification Handbook}}.
\newblock \bibinfo{publisher}{Logos Publishing House},
  \bibinfo{address}{Berlin, Germany}.
\newblock
\showISBNx{3832528199}


\bibitem[Hildebrandt et~al\mbox{.}(2016)]%
        {hildebrandt2016combining}
\bibfield{author}{\bibinfo{person}{Tobias Hildebrandt}, \bibinfo{person}{Felix
  Amerbauer}, {and} \bibinfo{person}{Stefanie Rinderle-Ma}.}
  \bibinfo{year}{2016}\natexlab{}.
\newblock \showarticletitle{Combining sonification and visualization for the
  analysis of process execution data}. In \bibinfo{booktitle}{\emph{Conference
  on Business Informatics}} (Paris, France) \emph{(\bibinfo{series}{CBI},
  Vol.~\bibinfo{volume}{02})}. \bibinfo{publisher}{IEEE Computer Society},
  \bibinfo{address}{Los Alamitos, CA, USA}, \bibinfo{pages}{32--37}.
\newblock
\urldef\tempurl%
\url{https://doi.org/10.1109/CBI.2016.47}
\showDOI{\tempurl}


\bibitem[Hoekstra et~al\mbox{.}(2014)]%
        {hoekstra2014robust}
\bibfield{author}{\bibinfo{person}{Rink Hoekstra}, \bibinfo{person}{Richard~D.
  Morey}, \bibinfo{person}{Jeffrey~N. Rouder}, {and} \bibinfo{person}{Eric-Jan
  Wagenmakers}.} \bibinfo{year}{2014}\natexlab{}.
\newblock \showarticletitle{Robust misinterpretation of confidence intervals}.
\newblock \bibinfo{journal}{\emph{Psychonomic Bulletin \& Review}}
  \bibinfo{volume}{21} (\bibinfo{year}{2014}), \bibinfo{pages}{1157--1164}.
\newblock
\urldef\tempurl%
\url{https://doi.org/10.3758/s13423-013-0572-3}
\showDOI{\tempurl}


\bibitem[Holloway et~al\mbox{.}(2022)]%
        {kimmarriott2022chi}
\bibfield{author}{\bibinfo{person}{Leona~M. Holloway}, \bibinfo{person}{Cagatay
  Goncu}, \bibinfo{person}{Alon Ilsar}, \bibinfo{person}{Matthew Butler}, {and}
  \bibinfo{person}{Kim Marriott}.} \bibinfo{year}{2022}\natexlab{}.
\newblock \showarticletitle{Infosonics: Accessible infographics for people who
  are blind using sonification and voice}. In
  \bibinfo{booktitle}{\emph{Proceedings of the 2022 CHI Conference on Human
  Factors in Computing Systems}} (New Orleans, LA, USA)
  \emph{(\bibinfo{series}{CHI})}. \bibinfo{publisher}{Association for Computing
  Machinery}, \bibinfo{address}{New York, NY, USA}, Article
  \bibinfo{articleno}{480}, \bibinfo{numpages}{13}~pages.
\newblock
\showISBNx{9781450391573}
\urldef\tempurl%
\url{https://doi.org/10.1145/3491102.3517465}
\showDOI{\tempurl}


\bibitem[Hoque et~al\mbox{.}(2023)]%
        {Hoque2023natural}
\bibfield{author}{\bibinfo{person}{Md~Naimul Hoque}, \bibinfo{person}{Md
  Ehtesham-Ul-Haque}, \bibinfo{person}{Niklas Elmqvist}, {and}
  \bibinfo{person}{Syed~Masum Billah}.} \bibinfo{year}{2023}\natexlab{}.
\newblock \showarticletitle{Accessible data representation with natural sound}.
  In \bibinfo{booktitle}{\emph{Proceedings of the 2023 CHI Conference on Human
  Factors in Computing Systems}} (Hamburg, Germany)
  \emph{(\bibinfo{series}{CHI})}. \bibinfo{publisher}{Association for Computing
  Machinery}, \bibinfo{address}{New York, NY, USA}, Article
  \bibinfo{articleno}{826}, \bibinfo{numpages}{19}~pages.
\newblock
\showISBNx{9781450394215}
\urldef\tempurl%
\url{https://doi.org/10.1145/3544548.3581087}
\showDOI{\tempurl}


\bibitem[Kamal et~al\mbox{.}(2021)]%
        {kamal2021recent}
\bibfield{author}{\bibinfo{person}{Aasim Kamal}, \bibinfo{person}{Parashar
  Dhakal}, \bibinfo{person}{Ahmad~Y. Javaid}, \bibinfo{person}{Vijay~K.
  Devabhaktuni}, \bibinfo{person}{Devinder Kaur}, \bibinfo{person}{Jack
  Zaientz}, {and} \bibinfo{person}{Robert Marinier}.}
  \bibinfo{year}{2021}\natexlab{}.
\newblock \showarticletitle{Recent advances and challenges in uncertainty
  visualization: a survey}.
\newblock \bibinfo{journal}{\emph{Journal of Visualization}}
  \bibinfo{volume}{24}, \bibinfo{number}{5} (\bibinfo{year}{2021}),
  \bibinfo{pages}{861--890}.
\newblock
\urldef\tempurl%
\url{https://doi.org/10.1007/s12650-021-00755-1}
\showDOI{\tempurl}


\bibitem[Kim et~al\mbox{.}(2024)]%
        {kim2024erie}
\bibfield{author}{\bibinfo{person}{Hyeok Kim}, \bibinfo{person}{Yea-Seul Kim},
  {and} \bibinfo{person}{Jessica Hullman}.} \bibinfo{year}{2024}\natexlab{}.
\newblock \showarticletitle{Erie: A declarative grammar for data sonification}.
  In \bibinfo{booktitle}{\emph{Proceedings of the 2024 CHI Conference on Human
  Factors in Computing Systems}} (Honolulu, HI, USA)
  \emph{(\bibinfo{series}{CHI})}. \bibinfo{publisher}{Association for Computing
  Machinery}, \bibinfo{address}{New York, NY, USA}, Article
  \bibinfo{articleno}{986}, \bibinfo{numpages}{19}~pages.
\newblock
\showISBNx{9798400703300}
\urldef\tempurl%
\url{https://doi.org/10.1145/3613904.3642442}
\showDOI{\tempurl}


\bibitem[Kramer(1994)]%
        {kramer1994intro}
\bibfield{author}{\bibinfo{person}{Gregory Kramer}.}
  \bibinfo{year}{1994}\natexlab{}.
\newblock \showarticletitle{An introduction to auditory display}.
\newblock In \bibinfo{booktitle}{\emph{Auditory Display: Sonification,
  Audification, and Auditory Interfaces}},
  \bibfield{editor}{\bibinfo{person}{Gregory Kramer}} (Ed.).
  \bibinfo{publisher}{Addison-Wesley}, \bibinfo{address}{Reading, MA, USA},
  \bibinfo{pages}{1--78}.
\newblock
\showISBNx{0201626047}


\bibitem[Kramer et~al\mbox{.}(1999)]%
        {kramer2010sonification}
\bibfield{author}{\bibinfo{person}{Gregory Kramer}, \bibinfo{person}{Bruce
  Walker}, \bibinfo{person}{Terri Bonebright}, \bibinfo{person}{Perry Cook},
  \bibinfo{person}{John~H. Flowers}, \bibinfo{person}{Nadine Miner}, {and}
  \bibinfo{person}{John Neuhoff}.} \bibinfo{year}{1999}\natexlab{}.
\newblock \bibinfo{booktitle}{\emph{Sonification report: Status of the field
  and research agenda}}.
\newblock \bibinfo{type}{{T}echnical {R}eport}. \bibinfo{institution}{Report
  prepared for the National Science Foundation by members of the International
  Community for Auditory Display}, \bibinfo{address}{Santa Fe, NM, USA}.
\newblock


\bibitem[Lacherez et~al\mbox{.}(2007)]%
        {lacherez2007overlapping}
\bibfield{author}{\bibinfo{person}{Philippe Lacherez}, \bibinfo{person}{Eunice
  Seah}, {and} \bibinfo{person}{Penelope Sanderson}.}
  \bibinfo{year}{2007}\natexlab{}.
\newblock \showarticletitle{Overlapping melodic alarms are almost
  indiscriminable}.
\newblock \bibinfo{journal}{\emph{Human Factors}} \bibinfo{volume}{49},
  \bibinfo{number}{4} (\bibinfo{year}{2007}), \bibinfo{pages}{637--645}.
\newblock
\urldef\tempurl%
\url{https://doi.org/10.1518/001872007X215719}
\showDOI{\tempurl}


\bibitem[Leitner and Buttenfield(2000)]%
        {leitner2000guidelines}
\bibfield{author}{\bibinfo{person}{Michael Leitner} {and}
  \bibinfo{person}{Barbara~P. Buttenfield}.} \bibinfo{year}{2000}\natexlab{}.
\newblock \showarticletitle{Guidelines for the display of attribute certainty}.
\newblock \bibinfo{journal}{\emph{Cartography and Geographic Information
  Science}} \bibinfo{volume}{27}, \bibinfo{number}{1} (\bibinfo{year}{2000}),
  \bibinfo{pages}{3--14}.
\newblock
\urldef\tempurl%
\url{https://doi.org/10.1559/152304000783548037}
\showDOI{\tempurl}


\bibitem[Lenzi et~al\mbox{.}(2021)]%
        {lenzi2021data}
\bibfield{author}{\bibinfo{person}{S. Lenzi}, \bibinfo{person}{P. Ciuccarelli},
  \bibinfo{person}{H. Liu}, {and} \bibinfo{person}{Y. Hua}.}
  \bibinfo{year}{2021}\natexlab{}.
\newblock \bibinfo{title}{Data sonification archive}.
\newblock
\newblock
\newblock
\shownote{\url{https://sonification.design/}, last access: 2024-03-17}.


\bibitem[Lotto and Holt(2011)]%
        {lotto2011psychology}
\bibfield{author}{\bibinfo{person}{Andrew Lotto} {and} \bibinfo{person}{Lori
  Holt}.} \bibinfo{year}{2011}\natexlab{}.
\newblock \showarticletitle{Psychology of auditory perception}.
\newblock \bibinfo{journal}{\emph{Wiley Interdisciplinary Reviews: Cognitive
  Science}} \bibinfo{volume}{2}, \bibinfo{number}{5} (\bibinfo{year}{2011}),
  \bibinfo{pages}{479--489}.
\newblock
\urldef\tempurl%
\url{https://doi.org/10.1002/wcs.123}
\showDOI{\tempurl}


\bibitem[MacEachren et~al\mbox{.}(2012)]%
        {maceachren2012visual}
\bibfield{author}{\bibinfo{person}{Alan~M. MacEachren},
  \bibinfo{person}{Robert~E. Roth}, \bibinfo{person}{James O'Brien},
  \bibinfo{person}{Bonan Li}, \bibinfo{person}{Derek Swingley}, {and}
  \bibinfo{person}{Mark Gahegan}.} \bibinfo{year}{2012}\natexlab{}.
\newblock \showarticletitle{Visual semiotics \& uncertainty visualization: An
  empirical study}.
\newblock \bibinfo{journal}{\emph{IEEE Transactions on Visualization and
  Computer Graphics}} \bibinfo{volume}{18}, \bibinfo{number}{12}
  (\bibinfo{year}{2012}), \bibinfo{pages}{2496--2505}.
\newblock
\urldef\tempurl%
\url{https://doi.org/10.1109/TVCG.2012.279}
\showDOI{\tempurl}


\bibitem[Marks(2004)]%
        {marks2004cross}
\bibfield{author}{\bibinfo{person}{Lawrence~E. Marks}.}
  \bibinfo{year}{2004}\natexlab{}.
\newblock \showarticletitle{Cross-modal interactions in speeded
  classification}.
\newblock In \bibinfo{booktitle}{\emph{The Handbook of Multisensory
  Processes}}, \bibfield{editor}{\bibinfo{person}{Gemma~A. Colvert},
  \bibinfo{person}{Charles Spence}, {and} \bibinfo{person}{Barry~E. Stein}}
  (Eds.). \bibinfo{publisher}{MIT Press}, \bibinfo{address}{Cambridge, MA,
  USA}, Chapter~6, \bibinfo{pages}{85--105}.
\newblock
\showISBNx{0-262-0331-6}
\urldef\tempurl%
\url{https://doi.org/10.7551/mitpress/3422.003.0009}
\showDOI{\tempurl}


\bibitem[Munzner(2015)]%
        {munzner2014visualization}
\bibfield{author}{\bibinfo{person}{Tamara Munzner}.}
  \bibinfo{year}{2015}\natexlab{}.
\newblock \bibinfo{booktitle}{\emph{Visualization Analysis and Design}}.
\newblock \bibinfo{publisher}{CRC Press}, \bibinfo{address}{Boca Raton, FL,
  USA}.
\newblock


\bibitem[Nees and Walker(2007)]%
        {nees2007listener}
\bibfield{author}{\bibinfo{person}{Michael~A. Nees} {and}
  \bibinfo{person}{Bruce~N. Walker}.} \bibinfo{year}{2007}\natexlab{}.
\newblock \showarticletitle{Listener, task, and auditory graph: Toward a
  conceptual model of auditory graph comprehension}. In
  \bibinfo{booktitle}{\emph{Proceedings of the International Conference on
  Auditory Display}} (Montréal, Canada) \emph{(\bibinfo{series}{ICAD})}.
  \bibinfo{pages}{266--273}.
\newblock


\bibitem[Neuhoff(2019)]%
        {neuhoff2019sonification}
\bibfield{author}{\bibinfo{person}{John~G. Neuhoff}.}
  \bibinfo{year}{2019}\natexlab{}.
\newblock \showarticletitle{Is sonification doomed to fail?}. In
  \bibinfo{booktitle}{\emph{International Conference on Auditory Display}}
  (Newcastle upon Tyne, United Kingdom) \emph{(\bibinfo{series}{ICAD})}.
  \bibinfo{pages}{327--330}.
\newblock
\urldef\tempurl%
\url{https://doi.org/10.21785/icad2019.069}
\showDOI{\tempurl}


\bibitem[Padilla et~al\mbox{.}(2022)]%
        {padilla2020uncertainty}
\bibfield{author}{\bibinfo{person}{Lace Padilla}, \bibinfo{person}{Matthew
  Kay}, {and} \bibinfo{person}{Jessica Hullman}.}
  \bibinfo{year}{2022}\natexlab{}.
\newblock \showarticletitle{Uncertainty visualization}.
\newblock In \bibinfo{booktitle}{\emph{Computational Statistics in Data
  Science}}, \bibfield{editor}{\bibinfo{person}{Walter~W. Piegorsch},
  \bibinfo{person}{Richard~A. Levine}, \bibinfo{person}{Hao~Helen Zhang}, {and}
  \bibinfo{person}{Thomas C.~M. Lee}} (Eds.). \bibinfo{publisher}{John Wiley \&
  Sons, Ltd.}, \bibinfo{address}{Hoboken, NJ, USA}, Chapter~21,
  \bibinfo{pages}{405--421}.
\newblock
\showISBNx{9781118445112}
\urldef\tempurl%
\url{https://doi.org/10.1002/9781118445112.stat08296}
\showDOI{\tempurl}


\bibitem[Parise and Spence(2013)]%
        {parise2013audiovisual}
\bibfield{author}{\bibinfo{person}{Cesare Parise} {and}
  \bibinfo{person}{Charles Spence}.} \bibinfo{year}{2013}\natexlab{}.
\newblock \showarticletitle{Audiovisual cross-modal correspondences in the
  general population}.
\newblock In \bibinfo{booktitle}{\emph{The Oxford Handbook of Synesthesia}},
  \bibfield{editor}{\bibinfo{person}{Julia Simner} {and}
  \bibinfo{person}{Edward Hubbard}} (Eds.). \bibinfo{publisher}{Oxford
  University Press}, \bibinfo{address}{Oxford, UK}, Chapter~39,
  \bibinfo{pages}{790--815}.
\newblock
\showISBNx{9780199603329}
\urldef\tempurl%
\url{https://doi.org/10.1093/oxfordhb/9780199603329.013.0039}
\showDOI{\tempurl}


\bibitem[Peres and Lane(2005)]%
        {peres2005auditory}
\bibfield{author}{\bibinfo{person}{S.~Camille Peres} {and}
  \bibinfo{person}{David~M. Lane}.} \bibinfo{year}{2005}\natexlab{}.
\newblock \showarticletitle{Auditory graphs: The effects of redundant
  dimensions and divided attention}. In \bibinfo{booktitle}{\emph{Proceedings
  of the International Conference on Auditory Display}} (Limerick, Ireland)
  \emph{(\bibinfo{series}{ICAD})}. \bibinfo{pages}{169--174}.
\newblock


\bibitem[Pinney et~al\mbox{.}(2023)]%
        {pinney2023aesthetic}
\bibfield{author}{\bibinfo{person}{Joel Pinney}, \bibinfo{person}{Fiona
  Carroll}, {and} \bibinfo{person}{Esyin Chew}.}
  \bibinfo{year}{2023}\natexlab{}.
\newblock \showarticletitle{Enhancing visual encodings of uncertainty through
  aesthetic depictions in line graph visualisations}. In
  \bibinfo{booktitle}{\emph{Human Interface and the Management of Information}}
  (Copenhagen, Denmark) \emph{(\bibinfo{series}{HCII 2023. Lecture Notes in
  Computer Science}, Vol.~\bibinfo{volume}{14015})}.
  \bibinfo{publisher}{Springer}, \bibinfo{address}{Cham, Switzerland},
  \bibinfo{pages}{272–291}.
\newblock
\showISBNx{978-3-031-35131-0}
\urldef\tempurl%
\url{https://doi.org/10.1007/978-3-031-35132-7_20}
\showDOI{\tempurl}


\bibitem[Portugal and Svaiter(2010)]%
        {portugal2011weber}
\bibfield{author}{\bibinfo{person}{R.~Doyle Portugal} {and}
  \bibinfo{person}{Benar~Fux Svaiter}.} \bibinfo{year}{2010}\natexlab{}.
\newblock \showarticletitle{Weber-Fechner law and the optimality of the
  logarithmic scale}.
\newblock \bibinfo{journal}{\emph{Minds and Machines}}  \bibinfo{volume}{21}
  (\bibinfo{year}{2010}), \bibinfo{pages}{73--81}.
\newblock
\urldef\tempurl%
\url{https://doi.org/10.1007/s11023-010-9221-z}
\showDOI{\tempurl}


\bibitem[Pringle-Pattison(1911)]%
        {pringle1911Weber}
\bibfield{author}{\bibinfo{person}{Andrew~Seth Pringle-Pattison}.}
  \bibinfo{year}{1911}\natexlab{}.
\newblock \showarticletitle{Weber's Law}.
\newblock In \bibinfo{booktitle}{\emph{Encyclop{\ae}dia Britannica}}.
  \bibinfo{publisher}{Cambridge University Press}, \bibinfo{pages}{458--459}.
\newblock


\bibitem[Robinson~Moore et~al\mbox{.}(2024)]%
        {Moore2024spatial}
\bibfield{author}{\bibinfo{person}{Wilfredo~Joshua Robinson~Moore},
  \bibinfo{person}{Medhani Kalal}, \bibinfo{person}{Jennifer~L. Tennison},
  \bibinfo{person}{Nicholas~A Giudice}, {and} \bibinfo{person}{Jenna
  Gorlewicz}.} \bibinfo{year}{2024}\natexlab{}.
\newblock \showarticletitle{Spatial audio-enhanced multimodal graph rendering
  for efficient data trend learning on touchscreen devices}. In
  \bibinfo{booktitle}{\emph{Proceedings of the CHI Conference on Human Factors
  in Computing Systems}} (Honolulu, HI, USA) \emph{(\bibinfo{series}{CHI})}.
  \bibinfo{publisher}{Association for Computing Machinery},
  \bibinfo{address}{New York, NY, USA}, Article \bibinfo{articleno}{206},
  \bibinfo{numpages}{12}~pages.
\newblock
\showISBNx{9798400703300}
\urldef\tempurl%
\url{https://doi.org/10.1145/3613904.3641959}
\showDOI{\tempurl}


\bibitem[R{\"o}nnberg(2019)]%
        {ronnberg2019sonification}
\bibfield{author}{\bibinfo{person}{Niklas R{\"o}nnberg}.}
  \bibinfo{year}{2019}\natexlab{}.
\newblock \showarticletitle{Sonification supports perception of brightness
  contrast}.
\newblock \bibinfo{journal}{\emph{Journal on Multimodal User Interfaces}}
  \bibinfo{volume}{13}, \bibinfo{number}{4} (\bibinfo{year}{2019}),
  \bibinfo{pages}{373--381}.
\newblock
\urldef\tempurl%
\url{https://doi.org/10.1007/s12193-019-00311-0}
\showDOI{\tempurl}


\bibitem[R{\"o}nnberg et~al\mbox{.}(2016)]%
        {ronnberg2016sonification}
\bibfield{author}{\bibinfo{person}{Niklas R{\"o}nnberg},
  \bibinfo{person}{Gustav Hallstr{\"o}m}, \bibinfo{person}{Tobias Erlandsson},
  {and} \bibinfo{person}{Jimmy Johansson}.} \bibinfo{year}{2016}\natexlab{}.
\newblock \showarticletitle{Sonification support for information visualization
  dense data displays}. In \bibinfo{booktitle}{\emph{VIS Infovis Posters}}
  (Baltimore, Maryland, USA). \bibinfo{publisher}{IEEE Computer Society},
  \bibinfo{address}{Los Alamitos, CA, USA}.
\newblock


\bibitem[Roth(2017)]%
        {roth2017visual}
\bibfield{author}{\bibinfo{person}{Robert~E. Roth}.}
  \bibinfo{year}{2017}\natexlab{}.
\newblock \showarticletitle{Visual variables}.
\newblock In \bibinfo{booktitle}{\emph{International Encyclopedia of Geography:
  People, the Earth, Environment and Technology}},
  \bibfield{editor}{\bibinfo{person}{Douglas Richardson}, \bibinfo{person}{Noel
  Castree}, \bibinfo{person}{Michael~F. Goodchild}, \bibinfo{person}{Audrey
  Kobayashi}, \bibinfo{person}{Weidong Liu}, {and} \bibinfo{person}{Richard~A.
  Marston}} (Eds.). \bibinfo{publisher}{John Wiley \& Sons, Ltd}.
\newblock
\showISBNx{9780470659632}
\urldef\tempurl%
\url{https://doi.org/10.1002/9781118786352.wbieg0761}
\showDOI{\tempurl}


\bibitem[Rubab et~al\mbox{.}(2023)]%
        {rubab2023exploring}
\bibfield{author}{\bibinfo{person}{Sadia Rubab}, \bibinfo{person}{Lingyun Yu},
  \bibinfo{person}{Junxiu Tang}, {and} \bibinfo{person}{Yingcai Wu}.}
  \bibinfo{year}{2023}\natexlab{}.
\newblock \showarticletitle{Exploring effective relationships between
  visual-audio channels in data visualization}.
\newblock \bibinfo{journal}{\emph{Journal of Visualization}}
  \bibinfo{volume}{26} (\bibinfo{year}{2023}), \bibinfo{pages}{937--956}.
\newblock
\urldef\tempurl%
\url{https://doi.org/10.1007/s12650-023-00909-3}
\showDOI{\tempurl}


\bibitem[Ruginski et~al\mbox{.}(2016)]%
        {ruginski2016non}
\bibfield{author}{\bibinfo{person}{Ian~T. Ruginski},
  \bibinfo{person}{Alexander~P. Boone}, \bibinfo{person}{Lace~M. Padilla},
  \bibinfo{person}{Le Liu}, \bibinfo{person}{Nahal Heydari},
  \bibinfo{person}{Heidi~S. Kramer}, \bibinfo{person}{Mary Hegarty},
  \bibinfo{person}{William~B. Thompson}, \bibinfo{person}{Donald~H. House},
  {and} \bibinfo{person}{Sarah~H. Creem-Regehr}.}
  \bibinfo{year}{2016}\natexlab{}.
\newblock \showarticletitle{Non-expert interpretations of hurricane forecast
  uncertainty visualizations}.
\newblock \bibinfo{journal}{\emph{Spatial Cognition \& Computation}}
  \bibinfo{volume}{16}, \bibinfo{number}{2} (\bibinfo{year}{2016}),
  \bibinfo{pages}{154--172}.
\newblock
\urldef\tempurl%
\url{https://doi.org/10.1080/13875868.2015.1137577}
\showDOI{\tempurl}


\bibitem[Sandor and Lane(2003)]%
        {sandor2003sonification}
\bibfield{author}{\bibinfo{person}{Aniko Sandor} {and} \bibinfo{person}{David
  Lane}.} \bibinfo{year}{2003}\natexlab{}.
\newblock \showarticletitle{Sonification of absolute values with single and
  multiple dimensions}. In \bibinfo{booktitle}{\emph{Proceedings of the 2003
  International Conference on Auditory Display}} (Boston, MA, USA)
  \emph{(\bibinfo{series}{ICAD})}. \bibinfo{publisher}{Georgia Institute of
  Technology}, \bibinfo{pages}{243--246}.
\newblock


\bibitem[Schloss et~al\mbox{.}(2019)]%
        {Schloss2019colormap}
\bibfield{author}{\bibinfo{person}{Karen~B. Schloss},
  \bibinfo{person}{Connor~C. Gramazio}, \bibinfo{person}{Allison~T. Silverman},
  \bibinfo{person}{Madeline~L. Parker}, {and} \bibinfo{person}{Audrey~S.
  Wang}.} \bibinfo{year}{2019}\natexlab{}.
\newblock \showarticletitle{Mapping color to meaning in colormap data
  visualizations}.
\newblock \bibinfo{journal}{\emph{IEEE Transactions on Visualization and
  Computer Graphics}} \bibinfo{volume}{25}, \bibinfo{number}{1}
  (\bibinfo{year}{2019}), \bibinfo{pages}{810--819}.
\newblock
\urldef\tempurl%
\url{https://doi.org/10.1109/TVCG.2018.2865147}
\showDOI{\tempurl}


\bibitem[Schoenlein et~al\mbox{.}(2023)]%
        {Schoenlein2023relational}
\bibfield{author}{\bibinfo{person}{Melissa~A. Schoenlein},
  \bibinfo{person}{Johnny Campos}, \bibinfo{person}{Kevin~J. Lande},
  \bibinfo{person}{Laurent Lessard}, {and} \bibinfo{person}{Karen~B. Schloss}.}
  \bibinfo{year}{2023}\natexlab{}.
\newblock \showarticletitle{Unifying effects of direct and relational
  associations for visual communication}.
\newblock \bibinfo{journal}{\emph{IEEE Transactions on Visualization and
  Computer Graphics}} \bibinfo{volume}{29}, \bibinfo{number}{1}
  (\bibinfo{year}{2023}), \bibinfo{pages}{385--395}.
\newblock
\urldef\tempurl%
\url{https://doi.org/10.1109/TVCG.2022.3209443}
\showDOI{\tempurl}


\bibitem[Schulz et~al\mbox{.}(2016)]%
        {schulz2016generative}
\bibfield{author}{\bibinfo{person}{Christoph Schulz}, \bibinfo{person}{Arlind
  Nocaj}, \bibinfo{person}{Mennatallah El{-}Assady}, \bibinfo{person}{Steffen
  Frey}, \bibinfo{person}{Marcel Hlawatsch}, \bibinfo{person}{Michael Hund},
  \bibinfo{person}{Grzegorz~Karol Karch}, \bibinfo{person}{Rudolf Netzel},
  \bibinfo{person}{Christin Sch{\"{a}}tzle}, \bibinfo{person}{Miriam Butt},
  \bibinfo{person}{Daniel~A. Keim}, \bibinfo{person}{Thomas Ertl},
  \bibinfo{person}{Ulrik Brandes}, {and} \bibinfo{person}{Daniel Weiskopf}.}
  \bibinfo{year}{2016}\natexlab{}.
\newblock \showarticletitle{Generative data models for validation and
  evaluation of visualization Techniques}. In
  \bibinfo{booktitle}{\emph{Proceedings of the Sixth Workshop on Beyond Time
  and Errors on Novel Evaluation Methods for Visualization}}
  \emph{(\bibinfo{series}{BELIV})}. \bibinfo{publisher}{Association for
  Computing Machinery}, \bibinfo{address}{New York, NY, USA},
  \bibinfo{pages}{112--124}.
\newblock
\urldef\tempurl%
\url{https://doi.org/10.1145/2993901.2993907}
\showDOI{\tempurl}


\bibitem[Snyder(2000)]%
        {snyder2000music}
\bibfield{author}{\bibinfo{person}{Bob Snyder}.}
  \bibinfo{year}{2000}\natexlab{}.
\newblock \bibinfo{booktitle}{\emph{Music and Memory: An Introduction}}.
\newblock \bibinfo{publisher}{MIT Press}, \bibinfo{address}{Cambridge, MA,
  USA}.
\newblock
\showISBNx{0-262-1944-4}


\bibitem[Thompson et~al\mbox{.}(2023)]%
        {bongshinlee2023chi}
\bibfield{author}{\bibinfo{person}{John~R. Thompson}, \bibinfo{person}{Jesse~J.
  Martinez}, \bibinfo{person}{Alper Sarikaya}, \bibinfo{person}{Edward
  Cutrell}, {and} \bibinfo{person}{Bongshin Lee}.}
  \bibinfo{year}{2023}\natexlab{}.
\newblock \showarticletitle{Chart reader: Accessible visualization experiences
  designed with screen reader users}. In \bibinfo{booktitle}{\emph{Proceedings
  of the 2023 CHI Conference on Human Factors in Computing Systems}} (Hamburg,
  Germany) \emph{(\bibinfo{series}{CHI})}. \bibinfo{publisher}{Association for
  Computing Machinery}, \bibinfo{address}{New York, NY, USA}, Article
  \bibinfo{articleno}{802}, \bibinfo{numpages}{18}~pages.
\newblock
\showISBNx{9781450394215}
\urldef\tempurl%
\url{https://doi.org/10.1145/3544548.3581186}
\showDOI{\tempurl}


\bibitem[Tomlinson et~al\mbox{.}(2020)]%
        {Tomlinson2020science}
\bibfield{author}{\bibinfo{person}{Brianna~J. Tomlinson},
  \bibinfo{person}{Bruce~N. Walker}, {and} \bibinfo{person}{Emily~B. Moore}.}
  \bibinfo{year}{2020}\natexlab{}.
\newblock \showarticletitle{Auditory display in interactive science
  simulations: Description and sonification support interaction and enhance
  opportunities for learning}. In \bibinfo{booktitle}{\emph{Proceedings of the
  2020 CHI Conference on Human Factors in Computing Systems}} (Honolulu, HI,
  USA) \emph{(\bibinfo{series}{CHI})}. \bibinfo{publisher}{Association for
  Computing Machinery}, \bibinfo{address}{New York, NY, USA}, Article
  \bibinfo{articleno}{757}, \bibinfo{numpages}{12}~pages.
\newblock
\showISBNx{9781450367080}
\urldef\tempurl%
\url{https://doi.org/10.1145/3313831.3376886}
\showDOI{\tempurl}


\bibitem[Vriend et~al\mbox{.}(2025)]%
        {Vriend:2025:DarusSupplemental}
\bibfield{author}{\bibinfo{person}{Sita Vriend}, \bibinfo{person}{David
  H{\"a}gele}, {and} \bibinfo{person}{Daniel Weiskopf}.}
  \bibinfo{year}{2025}\natexlab{}.
\newblock \bibinfo{title}{Supplemental materials for ``Two empirical studies on
  audiovisual semiotics of uncertainty''}.
\newblock
\newblock
\urldef\tempurl%
\url{https://doi.org/10.18419/DARUS-4137}
\showDOI{\tempurl}
\newblock
\shownote{DaRUS, version V1}.


\bibitem[Walker(2002)]%
        {walker2002magnitude}
\bibfield{author}{\bibinfo{person}{Bruce~N. Walker}.}
  \bibinfo{year}{2002}\natexlab{}.
\newblock \showarticletitle{Magnitude estimation of conceptual data dimensions
  for use in sonification}.
\newblock \bibinfo{journal}{\emph{Journal of Experimental Psychology: Applied}}
  \bibinfo{volume}{8}, \bibinfo{number}{4} (\bibinfo{year}{2002}),
  \bibinfo{pages}{211--221}.
\newblock
\urldef\tempurl%
\url{https://doi.org/10.1037//1076-898x.8.4.211}
\showDOI{\tempurl}


\bibitem[Walker(2007)]%
        {walker2007consistency}
\bibfield{author}{\bibinfo{person}{Bruce~N. Walker}.}
  \bibinfo{year}{2007}\natexlab{}.
\newblock \showarticletitle{Consistency of magnitude estimations with
  conceptual data dimensions used for sonification}.
\newblock \bibinfo{journal}{\emph{Applied Cognitive Psychology}}
  \bibinfo{volume}{21}, \bibinfo{number}{5} (\bibinfo{year}{2007}),
  \bibinfo{pages}{579--599}.
\newblock
\urldef\tempurl%
\url{https://doi.org/10.1002/acp.1291}
\showDOI{\tempurl}


\bibitem[Walker(2023)]%
        {Walker2023promise}
\bibfield{author}{\bibinfo{person}{Bruce~N. Walker}.}
  \bibinfo{year}{2023}\natexlab{}.
\newblock \showarticletitle{The past, present, and promise of sonification}.
\newblock \bibinfo{journal}{\emph{Arbor}} \bibinfo{volume}{199},
  \bibinfo{number}{810} (\bibinfo{year}{2023}), \bibinfo{pages}{a728}.
\newblock
\urldef\tempurl%
\url{https://doi.org/10.3989/arbor.2023.810008}
\showDOI{\tempurl}


\bibitem[Walker and Kramer(2005)]%
        {walker2005mappings}
\bibfield{author}{\bibinfo{person}{Bruce~N. Walker} {and}
  \bibinfo{person}{Gregory Kramer}.} \bibinfo{year}{2005}\natexlab{}.
\newblock \showarticletitle{Mappings and metaphors in auditory displays: An
  experimental assessment}.
\newblock \bibinfo{journal}{\emph{ACM Transactions on Applied Perception}}
  \bibinfo{volume}{2}, \bibinfo{number}{4} (\bibinfo{year}{2005}),
  \bibinfo{pages}{407–412}.
\newblock
\urldef\tempurl%
\url{https://doi.org/10.1145/1101530.1101534}
\showDOI{\tempurl}


\bibitem[Wang et~al\mbox{.}(2022)]%
        {wang2022seeing}
\bibfield{author}{\bibinfo{person}{Ruobin Wang}, \bibinfo{person}{Crescentia
  Jung}, {and} \bibinfo{person}{Y. Kim}.} \bibinfo{year}{2022}\natexlab{}.
\newblock \showarticletitle{Seeing through sounds: Mapping auditory dimensions
  to data and charts for people with visual impairments}.
\newblock \bibinfo{journal}{\emph{Computer Graphics Forum}}
  \bibinfo{volume}{41}, \bibinfo{number}{3} (\bibinfo{year}{2022}),
  \bibinfo{pages}{71--83}.
\newblock
\urldef\tempurl%
\url{https://doi.org/10.1111/cgf.14523}
\showDOI{\tempurl}


\bibitem[Weiskopf(2022)]%
        {Weiskopf2022Uncertainty}
\bibfield{author}{\bibinfo{person}{Daniel Weiskopf}.}
  \bibinfo{year}{2022}\natexlab{}.
\newblock \showarticletitle{Uncertainty visualization: Concepts, methods, and
  applications in biological data visualization}.
\newblock \bibinfo{journal}{\emph{Frontiers in Bioinformatics}}
  \bibinfo{volume}{2} (\bibinfo{year}{2022}).
\newblock
\showISSN{2673-7647}
\urldef\tempurl%
\url{https://doi.org/10.3389/fbinf.2022.793819}
\showDOI{\tempurl}


\bibitem[Witt et~al\mbox{.}(2023)]%
        {witt2023visualizing}
\bibfield{author}{\bibinfo{person}{Jessica Witt}, \bibinfo{person}{Zachary
  Labe}, \bibinfo{person}{Amelia Warden}, {and} \bibinfo{person}{Benjamin
  Clegg}.} \bibinfo{year}{2023}\natexlab{}.
\newblock \showarticletitle{Visualizing uncertainty in hurricane forecasts with
  animated risk trajectories}.
\newblock \bibinfo{journal}{\emph{Weather, Climate, and Society}}
  \bibinfo{volume}{15} (\bibinfo{year}{2023}), \bibinfo{pages}{407--424}.
\newblock
\urldef\tempurl%
\url{https://doi.org/10.1175/WCAS-D-21-0173.1}
\showDOI{\tempurl}


\bibitem[Yang et~al\mbox{.}(2024)]%
        {yang2024trust}
\bibfield{author}{\bibinfo{person}{Fumeng Yang}, \bibinfo{person}{Chloe~Rose
  Mortenson}, \bibinfo{person}{Erik Nisbet}, \bibinfo{person}{Nicholas
  Diakopoulos}, {and} \bibinfo{person}{Matthew Kay}.}
  \bibinfo{year}{2024}\natexlab{}.
\newblock \showarticletitle{In dice we trust: Uncertainty displays for
  maintaining trust in election forecasts over time}. In
  \bibinfo{booktitle}{\emph{Proceedings of the CHI Conference on Human Factors
  in Computing Systems}} (Honolulu, HI, USA) \emph{(\bibinfo{series}{CHI})}.
  \bibinfo{publisher}{Association for Computing Machinery},
  \bibinfo{address}{New York, NY, USA}, Article \bibinfo{articleno}{389},
  \bibinfo{numpages}{24}~pages.
\newblock
\showISBNx{9798400703300}
\urldef\tempurl%
\url{https://doi.org/10.1145/3613904.3642371}
\showDOI{\tempurl}


\end{thebibliography}

\newpage
\appendix

\section{Compare Color Task Results} \label{clr_task}

\begin{table}[h]
    \caption{Accuracy proportion of the image comparison across participants before exclusion. The image names under image 1 and 2 refer to the file names of the relevant stimuli.}
    \label{tab:img_comparison}
    \centering
    \begin{tabular}{lll} \toprule  
         Image 1&  Image 2&  Accuracy\\\midrule
         brightness\_0&  brightness\_0&  0.9539171\\
         brightness\_0&  brightness\_1&  0.9861751\\\midrule 
         brightness\_1&  brightness\_1&  0.8801843\\
         brightness\_1&  brightness\_2&  0.9677419\\\midrule 
 brightness\_2& brightness\_0&0.9861751\\ 
 brightness\_2& brightness\_2&0.8064516\\\midrule 
 saturation\_0& saturation\_0&0.8847926\\
 saturation\_0& saturation\_1&0.9953917\\\midrule 
 saturation\_1& saturation\_1&0.8156682\\ 
 saturation\_1& saturation\_2&0.9585253\\\midrule 
 saturation\_2& saturation\_0&0.9907834\\
 saturation\_2& saturation\_2&0.8110599\\\bottomrule
    \end{tabular}
\end{table}

\section{Compare Sound Task Results} \label{snd_task}

\begin{table}[h]
    \caption{Accuracy proportion of the sound comparison across participants before exclusion. The sound names under sound 1 and 2 refer to the file names of the relevant stimuli. The pitch stimuli are the files named 'frequency'.}
    \label{tab:snd_comparison}
    \centering
    \begin{tabular}{lll}  \toprule  
         Sound 1&  Sound 2& Accuracy\\\midrule
         frequency\_240&  frequency\_440& 0.9861751\\ 
         frequency 240&  frequency\_640& 1.0000000\\ 
         frequency\_240&  frequency\_240& 0.9354839\\
         frequency\_640&  frequency\_640& 0.9447005\\\midrule  
         loudness\_1&  loudness\_0& 0.9262673\\
         loudness\_1&  loudness\_1& 0.9308756\\
         loudness\_2&  loudness\_1& 0.6820276\\
         loudness\_2&  loudness\_2& 0.9216590\\\bottomrule
    \end{tabular}
\end{table}

\newpage

\section{Experiment 1: Preference Consistency} \label{exp1_consistency}

\begin{table}[h]
    \centering
    \caption{The proportion of consistent visual polarity preferences before exclusion.}
    \begin{tabular}[h!]{lll} \toprule 
        Audio condition&  Visual condition& Consistency\\ \midrule 
         loudness&  brightness& 0.7373272\\
         loudness&  fuzzy& 0.8479263\\ 
         loudness&  saturation& 0.8525346\\ 
         loudness&  size& 0.8525346\\ \midrule  
         pitch&  brightness& 0.8479263\\ 
 pitch& fuzzy&0.8110599\\ 
 pitch& saturation&0.8248848\\  
 pitch& size&0.7880184\\ \midrule  
 tempo& brightness&0.7511521\\  
 tempo& fuzzy&0.7511521\\  
 tempo& saturation&0.7373272\\  
 tempo& size&0.7880184\\ \bottomrule
    \end{tabular}
    \label{tab:consistency_all}
\end{table}

\begin{table}[h]
    \centering
    \caption{The proportion of consistent visual polarity preferences of the audio condition before exclusion.}
    \begin{tabular}[h!]{ll} \toprule 
         Audio condition& Consistency\\ \midrule 
         loudness& 0.8225806\\
         pitch& 0.8179724\\
         tempo& 0.7569124\\ \bottomrule
    \end{tabular}
    \label{tab:concistency_snd}
\end{table}

\begin{table}[!t]
    \centering
    \caption{The proportion of consistent visual polarity preferences of the visual condition before exclusion.}
    \begin{tabular}[h!]{ll} \toprule 
         Visual condition& Consistency\\ \midrule 
         brightness& 0.7788018\\ 
         fuzzy& 0.8033794\\ 
         saturation& 0.8049155\\ 
         size& 0.8095238\\ \bottomrule
    \end{tabular}
    \label{tab:consistency_vis}
\end{table}

\clearpage

\onecolumn
\section{Experiment 1: Exact Binomial Test Results} \label{exp1_mapping}

\begin{table}[!h]
    \caption{The results of the exact binomial test on the polarity preference per AV stimulus.}
    \label{fig:exp1_binom_table}
    \centering
    \begin{tabular}{lllllll}    \toprule
        Visual condition & Audio condition & Visual polarity pref. & Binomial prop. & 95\% CI& p-value \\ \midrule 
        brightness & pitch & negative & 0.140 & 0.168, 0.115 & 0.000 \\ 
        brightness & pitch & positive & 0.860 & 0.885, 0.832 & 0.000 \\ 
        brightness & loudness & negative & 0.426 & 0.465, 0.389 & 0.000 \\ 
        brightness & loudness & positive & 0.574 & 0.611, 0.535 & 0.000 \\ 
        brightness & tempo & negative & 0.353 & 0.390, 0.317 & 0.000 \\ 
        brightness & tempo & positive & 0.647 & 0.683, 0.610 & 0.000 \\ \midrule
        fuzzy & pitch & negative & 0.725 & 0.758, 0.690 & 0.000 \\ 
        fuzzy & pitch & positive & 0.275 & 0.310, 0.242 & 0.000 \\ 
        fuzzy & loudness & negative & 0.831 & 0.858, 0.801 & 0.000 \\ 
        fuzzy & loudness & positive & 0.169 & 0.199, 0.142 & 0.000 \\ 
        fuzzy & tempo & negative & 0.721 & 0.754, 0.685 & 0.000 \\ 
        fuzzy & tempo & positive & 0.279 & 0.315, 0.246 & 0.000 \\ \midrule
        saturation & pitch & negative & 0.215 & 0.248, 0.184 & 0.000 \\ 
        saturation & pitch & positive & 0.785 & 0.816, 0.752 & 0.000 \\ 
        saturation & loudness & negative & 0.160 & 0.190, 0.134 & 0.000 \\ 
        saturation & loudness & positive & 0.840 & 0.866, 0.810 & 0.000 \\ 
        saturation & tempo & negative & 0.256 & 0.290, 0.223 & 0.000 \\ 
        saturation & tempo & positive & 0.744 & 0.777, 0.710 & 0.000 \\ \midrule
        size & pitch & negative & 0.599 & 0.636, 0.561 & 0.000 \\ 
        size & pitch & positive & 0.401 & 0.439, 0.364 & 0.000 \\ 
        size & loudness & negative & 0.159 & 0.189, 0.132 & 0.000 \\ 
        size & loudness & positive & 0.841 & 0.868, 0.811 & 0.000 \\ 
        size & tempo & negative & 0.594 & 0.631, 0.556 & 0.000 \\ 
        size & tempo & positive & 0.406 & 0.444, 0.369 & 0.000 \\ \bottomrule
    \end{tabular}
\end{table}

\clearpage

\section{Reaction Time Pairwise Analysis Results}  \label{exp1_rt_pair}

\begin{table}[!h]
    \centering
    \caption{The results of the posthoc pairwise analysis. }
    \begin{tabular}{llllll} \toprule
        Audio condition & Visual conditions & Statistic & Estimate & 95\% CI & p.adj \\  \midrule
        pitch & brightness-fuzzy & -4.737 & -333.900 & -473.061, -194.739 & 0.000 \\ 
        pitch & brightness-saturation & -3.411 & -262.351 & -414.193, -110.510 & 0.005 \\ 
        pitch & brightness-size & -3.196 & -266.203 & -430.609, -101.797 & 0.010 \\ 
        pitch & fuzzy-saturation & 0.952 & 71.549 & -76.762, 219.859 & 1.000 \\ 
        pitch & fuzzy-size & 0.735 & 67.697 & -114.115, 249.509 & 1.000 \\ 
        pitch & saturation-size & -0.045 & -3.851 & -172.644, 164.941 & 1.000 \\ \midrule
        loudness & brightness-fuzzy & 3.584 & 303.802 & 136.459, 471.145 & 0.003 \\ 
        loudness & brightness-saturation & 4.381 & 363.565 & 199.728, 527.403 & 0.000 \\ 
        loudness & brightness-size & 5.161 & 454.304 & 280.540, 628.068 & 0.000 \\ 
        loudness & fuzzy-saturation & 0.876 & 59.763 & -74.853, 194.379 & 1.000 \\ 
        loudness & fuzzy-size & 2.181 & 150.502 & 14.252, 286.752 & 0.184 \\ 
        loudness & saturation-size & 1.309 & 90.739 & -46.129, 227.607 & 1.000 \\ \midrule
        tempo & brightness-fuzzy & -0.794 & -61.910 & -215.774, 91.953 & 1.000 \\ 
        tempo & brightness-saturation & -0.262 & -18.994 & -162.202, 124.214 & 1.000 \\ 
        tempo & brightness-size & 0.856 & 61.548 & -80.328, 203.423 & 1.000 \\ 
        tempo & fuzzy-saturation & 0.532 & 42.916 & -116.458, 202.291 & 1.000 \\ 
        tempo & fuzzy-size & 1.460 & 123.458 & -43.451, 290.368 & 0.876 \\ 
        tempo & saturation-size & 1.021 & 80.542 & -75.238, 236.322 & 1.000 \\ \bottomrule
    \end{tabular}
    \label{fig:exp1_pair_table}
\end{table}

\clearpage

\section{Experiment 2: Exact Binomial Test Results} \label{exp2_mapping}

\begin{longtable}[c]{llllllll}
\caption{The results of the exact binomial test on the probability polarity preference per AV stimulus.\label{long}}\\

 \toprule
 Audio condition & Audio pol. pref. & Visual condition & Visual pol. pref. & Binomial prob. & 95\% CI & p-value \\
 \midrule
 \endfirsthead
 
 \multicolumn{8}{l}{Continuation of Table \ref{long}}\\
 \toprule
 Audio condition & Audio pol. pref. & Visual condition & Visual pol. pref. & Binomial prob. & 95\% CI & p-value \\
\midrule
 \endhead

 %\endfoot
 \bottomrule
 \endlastfoot
 
pitch & negative & brightness & negative & 0.181 & 0.154, 0.211 & 0.000 \\ 
        pitch & positive & brightness & negative & 0.277 & 0.245, 0.311 & 0.090 \\ 
        pitch & negative & brightness & positive & 0.136 & 0.113, 0.163 & 0.000 \\ 
        pitch & positive & brightness & positive & 0.405 & 0.370, 0.442 & 0.000 \\ \midrule 
        pitch & negative & fuzzy & negative & 0.207 & 0.178, 0.238 & 0.007 \\ 
        pitch & positive & fuzzy & negative & 0.514 & 0.477, 0.550 & 0.000 \\ 
        pitch & negative & fuzzy & positive & 0.116 & 0.094, 0.142 & 0.000 \\ 
        pitch & positive & fuzzy & positive & 0.164 & 0.138, 0.192 & 0.000 \\ \midrule
        pitch & negative & saturation & negative & 0.124 & 0.101, 0.150 & 0.000 \\ 
        pitch & positive & saturation & negative & 0.120 & 0.098, 0.146 & 0.000 \\ 
        pitch & negative & saturation & positive & 0.180 & 0.153, 0.209 & 0.000 \\ 
        pitch & positive & saturation & positive & 0.576 & 0.539, 0.612 & 0.000 \\ \midrule
        pitch & negative & size & negative & 0.086 & 0.067, 0.109 & 0.000 \\ 
        pitch & positive & size & negative & 0.155 & 0.130, 0.184 & 0.000 \\ 
        pitch & negative & size & positive & 0.207 & 0.178, 0.238 & 0.007 \\ 
        pitch & positive & size & positive & 0.551 & 0.515, 0.588 & 0.000 \\ \midrule
        loudness & negative & brightness & negative & 0.131 & 0.108, 0.158 & 0.000 \\ 
        loudness & positive & brightness & negative & 0.326 & 0.292, 0.361 & 0.000 \\ 
        loudness & negative & brightness & positive & 0.130 & 0.106, 0.156 & 0.000 \\ 
        loudness & positive & brightness & positive & 0.414 & 0.378, 0.450 & 0.000 \\ \midrule
        loudness & negative & fuzzy & negative & 0.135 & 0.111, 0.162 & 0.000 \\ 
        loudness & positive & fuzzy & negative & 0.539 & 0.503, 0.576 & 0.000 \\ 
        loudness & negative & fuzzy & positive & 0.151 & 0.126, 0.179 & 0.000 \\ 
        loudness & positive & fuzzy & positive & 0.174 & 0.148, 0.204 & 0.000 \\ \midrule
        loudness & negative & saturation & negative & 0.130 & 0.106, 0.156 & 0.000 \\ 
        loudness & positive & saturation & negative & 0.104 & 0.083, 0.128 & 0.000 \\ 
        loudness & negative & saturation & positive & 0.162 & 0.136, 0.191 & 0.000 \\ 
        loudness & positive & saturation & positive & 0.604 & 0.568, 0.639 & 0.000 \\ \midrule
        loudness & negative & size & negative & 0.112 & 0.090, 0.137 & 0.000 \\ 
        loudness & positive & size & negative & 0.147 & 0.123, 0.175 & 0.000 \\ 
        loudness & negative & size & positive & 0.138 & 0.114, 0.165 & 0.000 \\ 
        loudness & positive & size & positive & 0.603 & 0.566, 0.638 & 0.000 \\ \midrule
        tempo & negative & brightness & negative & 0.134 & 0.110, 0.160 & 0.000 \\ 
        tempo & positive & brightness & negative & 0.288 & 0.255, 0.322 & 0.019 \\ 
        tempo & negative & brightness & positive & 0.158 & 0.133, 0.186 & 0.000 \\ 
        tempo & positive & brightness & positive & 0.420 & 0.384, 0.457 & 0.000 \\ \midrule
        tempo & negative & fuzzy & negative & 0.186 & 0.159, 0.216 & 0.000 \\ 
        tempo & positive & fuzzy & negative & 0.515 & 0.478, 0.551 & 0.000 \\ 
        tempo & negative & fuzzy & positive & 0.127 & 0.104, 0.153 & 0.000 \\ 
        tempo & positive & fuzzy & positive & 0.172 & 0.145, 0.201 & 0.000 \\ \midrule
        tempo & negative & saturation & negative & 0.109 & 0.088, 0.134 & 0.000 \\ 
        tempo & positive & saturation & negative & 0.123 & 0.100, 0.149 & 0.000 \\ 
        tempo & negative & saturation & positive & 0.200 & 0.172, 0.231 & 0.001 \\ 
        tempo & positive & saturation & positive & 0.568 & 0.531, 0.604 & 0.000 \\ \midrule
        tempo & negative & size & negative & 0.118 & 0.095, 0.143 & 0.000 \\ 
        tempo & positive & size & negative & 0.169 & 0.143, 0.198 & 0.000 \\ 
        tempo & negative & size & positive & 0.172 & 0.145, 0.201 & 0.000 \\ 
        tempo & positive & size & positive & 0.542 & 0.505, 0.578 & 0.000 
    \label{fig:exp2_binom_table}
    \end{longtable}

\clearpage

\section{Experiment 2: Reaction Time Pairwise Analysis}

\begin{table}[!h]
    \centering
    \caption{The results of the posthoc pairwise analysis. }
    \begin{tabular}{llllllll} \toprule
        \textbf{Audio condition} & \textbf{Visual conditions} & \textbf{Polarity mapping} & \textbf{Statistic} & \textbf{Estimate} & \textbf{95\% CI} & \textbf{p.adj} \\ \midrule 
        pitch & brightness-fuzzy & matching & 3E-02 & 2E-07 & -1E-05, 1E-05 & 1.000 \\ 
        pitch & brightness-saturation & matching & -7E-01 & -5E-06 & -2E-05, 1E-05 & 1.000 \\ 
        pitch & brightness-size & matching & -5E-01 & -4E-06 & -2E-05, 1E-05 & 1.000 \\ 
        pitch & fuzzy-saturation & matching & -7E-01 & -5E-06 & -2E-05, 9E-06 & 1.000 \\ 
        pitch & fuzzy-size & matching & -6E-01 & -4E-06 & -2E-05, 9E-06 & 1.000 \\ 
        pitch & saturation-size & matching & 2E-01 & 1E-06 & -1E-05, 2E-05 & 1.000 \\ \midrule
        pitch & brightness-fuzzy & diverging & -2E+00 & -1E-05 & -3E-05, -1E-06 & 0.169 \\ 
        pitch & brightness-saturation & diverging & -1E+00 & -6E-06 & -2E-05, 6E-06 & 1.000 \\ 
        pitch & brightness-size & diverging & 1E+00 & 8E-06 & -4E-06, 2E-05 & 1.000 \\ 
        pitch & fuzzy-saturation & diverging & 1E+00 & 7E-06 & -6E-06, 2E-05 & 1.000 \\ 
        pitch & fuzzy-size & diverging & 3E+00 & 2E-05 & 9E-06, 3E-05 & 0.004 \\ 
        pitch & saturation-size & diverging & 2E+00 & 1E-05 & 2E-06, 3E-05 & 0.169 \\ \midrule
        loudness & brightness-fuzzy & matching & 2E+00 & 1E-05 & -2E-06, 3E-05 & 0.577 \\ 
        loudness & brightness-saturation & matching & -9E-01 & -6E-06 & -2E-05, 7E-06 & 1.000 \\ 
        loudness & brightness-size & matching & -2E+00 & -2E-05 & -3E-05, -2E-06 & 0.139 \\ 
        loudness & fuzzy-saturation & matching & -2E+00 & -2E-05 & -3E-05, -3E-06 & 0.118 \\ 
        loudness & fuzzy-size & matching & -4E+00 & -3E-05 & -4E-05, -1E-05 & 0.001 \\ 
        loudness & saturation-size & matching & -1E+00 & -9E-06 & -2E-05, 4E-06 & 1.000 \\ \midrule 
        loudness & brightness-fuzzy & diverging & -2E+00 & -1E-05 & -2E-05, 3E-06 & 0.672 \\ 
        loudness & brightness-saturation & diverging & 1E+00 & 8E-06 & -5E-06, 2E-05 & 1.000 \\ 
        loudness & brightness-size & diverging & 2E+00 & 1E-05 & -8E-07, 2E-05 & 0.391 \\ 
        loudness & fuzzy-saturation & diverging & 3E+00 & 2E-05 & 6E-06, 3E-05 & 0.026 \\ 
        loudness & fuzzy-size & diverging & 4E+00 & 2E-05 & 1E-05, 4E-05 & 0.003 \\ 
        loudness & saturation-size & diverging & 6E-01 & 4E-06 & -9E-06, 2E-05 & 1.000 \\ \midrule
        tempo & brightness-fuzzy & matching & 2E+00 & 1E-05 & -1E-06, 2E-05 & 0.497 \\ 
        tempo & brightness-saturation & matching & -1E+00 & -8E-06 & -2E-05, 4E-06 & 1.000 \\ 
        tempo & brightness-size & matching & -4E-01 & -3E-06 & -2E-05, 1E-05 & 1.000 \\ 
        tempo & fuzzy-saturation & matching & -3E+00 & -2E-05 & -3E-05, -7E-06 & 0.014 \\ 
        tempo & fuzzy-size & matching & -2E+00 & -1E-05 & -3E-05, -2E-06 & 0.128 \\ 
        tempo & saturation-size & matching & 8E-01 & 5E-06 & -7E-06, 2E-05 & 1.000 \\ \midrule
        tempo & brightness-fuzzy & diverging & -2E-01 & -9E-07 & -1E-05, 1E-05 & 1.000 \\ 
        tempo & brightness-saturation & diverging & 2E+00 & 1E-05 & 1E-07, 2E-05 & 0.290 \\ 
        tempo & brightness-size & diverging & 2E+00 & 1E-05 & -2E-06, 2E-05 & 0.624 \\ 
        tempo & fuzzy-saturation & diverging & 2E+00 & 1E-05 & -3E-07, 3E-05 & 0.334 \\ 
        tempo & fuzzy-size & diverging & 2E+00 & 1E-05 & -1E-06, 2E-05 & 0.478 \\ 
        tempo & saturation-size & diverging & -2E-01 & -2E-06 & -1E-05, 1E-05 & 1.000 \\ \bottomrule
    \end{tabular}
    \label{fig:exp2_pair_table}
\end{table}

\end{document}